\documentclass[aps,prd,10pt,superscriptaddress,groupedaddress,showpacs]{revtex4-1} 
\def\refitem#1{\relax} 

\usepackage{multirow}
\usepackage{hyperref}
\usepackage{graphicx}
\usepackage{amsfonts,amsmath,amssymb,bm,bbm,amssymb}
\usepackage{color}
\usepackage{colortbl}
\usepackage{dcolumn}
\usepackage{enumitem}
\usepackage{textcomp}
\usepackage{mathrsfs}

\newcommand{\tabincell}[2]{\begin{tabular}{@{}#1@{}}#2\end{tabular}}

\newcommand{\eq}{Eq.}

\newcommand{\fig}{Fig.}

\newcommand{\Ref}{Ref.}

\newcommand{\equ}[1]{\eq~(\ref{equ:#1})}
\newcommand{\figu}[1]{\fig~\ref{fig:#1}}

\hypersetup{
    pdfnewwindow=true,      % links in new window
    colorlinks=true,       % false: boxed links; true: colored links
    linkcolor=blue,          % color of internal links
    citecolor=blue,        % color of links to bibliography
    filecolor=blue,      % color of file links
    urlcolor=blue        % color of external links
}

\def\bi{\begin{itemize}[noitemsep,leftmargin=*]}
\def\ei{\end{itemize}}

\definecolor{mygray}{gray}{.9}
\definecolor{mypink}{rgb}{.99,.91,.95}
\definecolor{green(html/cssgreen)}{rgb}{0.0, 0.5, 0.0}
\definecolor{cyan}{rgb}{0.0, 1.0, 1.0}
\definecolor{pink}{rgb}{1.0, 0.75, 0.8}

\begin{document}

\title{Study of Non-Standard Charged-Current Interactions at the MOMENT experiment}
\author{Jian Tang}
\email[]{tangjian5@mail.sysu.edu.cn}
\author{Yibing Zhang}
\email[]{zhangyb27@mail2.sysu.edu.cn}
\affiliation{School of Physics, Sun Yat-Sen University, 510275, Guangzhou, China}

\date{\today}

\begin{abstract}
MuOn-decay MEdium baseline NeuTrino beam experiment (MOMENT) is a next-generation accelerator neutrino experiment, which can be used to probe new physics beyond Standard Model. We try to simulate neutrino oscillations confronting with Charged-Current and Non-Standard neutrino Interactions(CC-NSIs) at MOMENT. These NSIs could alter neutrino production and detection processes and interfere with neutrino oscillation channels. We separate a perturbative discussion of oscillation channels at near and far detectors, and analyze parameter correlations with the impact of CC-NSIs. Taking $\delta_{cp}$ and $\theta_{23}$ as an example, we find that CC-NSIs can induce bias in precision measurements of standard oscillation parameters. In addition, a combination of near and far detectors using Gd-doped water cherenkov technology at MOMENT is able to provide good constraints of CC-NSIs happening to the neutrino production and detection processes.\\

\end{abstract}


\maketitle
%\tableofcontents
\section{Introduction}
\label{sec:introduction}
In the past decades, we have seen enormous progress from neutrino oscillation experiments using solar, atmospheric, accelerator and reactor neutrinos~\cite{Aharmim:2011vm,Wendell:2010md,Abe:2008aa,Abe:2014ugx,Adamson:2016xxw,Abe:2014bwa,An:2016ses,RENO:2015ksa}. 
In the framework of three neutrino oscillations, there are six physics parameters
including three mixing angles $\theta_{12}$, $\theta_{13}$, $\theta_{23}$, one Dirac CP phase $\delta_{cp}$ and two mass squared splittings $\Delta m^2_{31}$, $\Delta m^2_{21}$. According to a global analysis of these neutrino oscillation experiments~\cite{Fogli:2012ua,Forero:2014bxa,Gonzalez-Garcia:2014bfa,Esteban:2016qun}, mixing angles $\theta_{12}$, $\theta_{13}$ \& $\theta_{23}$ and mass square differences $\Delta m^2_{21}$ \& $|\Delta m_{31}^2|$ have so far been well measured. 
The mixing angle $\theta_{23}$, however, has not been determined with enough precision to disentangle whether the mixing angle $\theta_{23}$ is $45^\circ$, while many discrete models point to a maximal mixing $\theta_{23}=45^\circ$ with regard to a $\mu-\tau$ symmetry. 
In addition, a deviation from $\theta_{23}=45^\circ$ causes an octant degeneracy problem in certain neutrino oscillation channels~\cite{Barger:2002rr,Minakata:2002qi}. 
Nonetheless, the Dirac CP phase describing the difference between matter and anti-matter as well as the sign of $\Delta m^2_{31}$ (normal mass hierarchy:$\Delta m^2_{31}>0$; inverted mass hierarchy: $\Delta m^2_{31}<0$) have not been well constrained yet. Though recent results from T2K~\cite{Abe:2015awa} and NO$\nu$A~\cite{Adamson:2016tbq} disfavor the inverted mass hierarchy at a low confidence level and give hints of $\delta_{CP}\approx -90^\circ$, we expect more data to draw a solid conclusion or further call for the next-generation experiments such as accelerator neutrino oscillation experiments like DUNE~\cite{Acciarri:2015uup} and T2HK~\cite{Abe:2014oxa}, the medium-baseline reactor experiments~\cite{Cao:2017hno} like JUNO~\cite{An:2015jdp} and RENO-50~\cite{Kim:2014rfa}, atmospheric neutrino experiments like INO~\cite{Kumar:2017sdq}, PINGU~\cite{Aartsen:2014oha} and KM3Net~\cite{Adrian-Martinez:2016fdl}.

Neutrinos are massless in Standard Model (SM) and phenomenon of neutrino oscillations is new physics beyond SM. 
It is required to generate massive neutrinos by extending SM, such popular candidates as seesaw models, supersymmetry models, extra-dimension models and the like. With more particle contents in new physics models, it might contain the sub-leading effects induced by non-standard neutrino interactions (NSIs) in nature. Effective operators towards this direction have been adopted to link neutrino mass models and NSIs~\cite{Gavela:2008ra,Bonnet:2009ej,Krauss:2011ur,Bonnet:2012kz}. Though there are viable models for sizable NSIs associated with neutral-current interactions~\cite{Farzan:2015doa, Farzan:2015hkd, Farzan:2016wym, Forero:2016ghr, Farzan:2016fmy}, it is still a tough task to model CC-NSIs surviving the overwhelming constraints from precision measurements of charged lepton properties.
Several studies have been conducted on NSIs from the experimental and model-building point of view~\cite{GonzalezGarcia:1998hj, Bergmann:1997mr, Bergmann:1998rg, Bergmann:2000gp, Bergmann:2000gn, Gago:2001xg, Gago:2001si, Guzzo:2004ue, Esmaili:2013fva, Li:2014mlo, Liao:2016hsa, Liao:2016reh, Liao:2016bgf, Salvado:2016uqu, Liao:2016orc, Coloma:2017egw, Datta:2017pfz, Liao:2017awz}. 
% \textcolor{red}{References about NC-NSIs suggested by the refereee are added here!} 
A review of NSIs is given in detail in \Ref~\cite{Ohlsson:2012kf,Miranda:2015dra}. With the help of an effective field theory, we can generally integrate out the mediator/propagator in the Feynman diagram and keep four fermions contact with each other. New physics scale is then embedded into the effective coupling constant $\epsilon_{\alpha^\prime \beta^\prime}^{\alpha\beta}$ where $\alpha/\beta$ or $\alpha^\prime \beta^\prime$ are the related fermion flavours. In theory the higher the new physics scale, the harder it is to reach a small effective coupling constant. We have reached an era of precision measurements of neutrino mixing parameters after an establishment of neutrino oscillation. It is promising for us to develop better neutrino detectors to search for sub-leading NSIs in the current and next-generation neutrino oscillation experiments as a complementary to the new physics search with the high intensity machine at the collider. 

The MuOn-decay MEdium baseline NeuTrino beam experiment (MOMENT) is a next-generation accelerator neutrino experiment proposed for discovery of leptonic CP violation~\cite{Cao:2014bea}. The atmospheric neutrino flux is a serious hindrance to the study of CP violation at MOMENT. Backgrounds caused by the atmospheric neutrinos exceeds the oscillation signal events significantly at O(100) MeV. Neutrino beams from such a continuous proton accelerator provide high luminosity fluxes but result in a loss of timing information which is traditionally used to suppress atmospheric neutrino backgrounds in the accelerator neutrino oscillation experiment with the pulsed proton beam facility. A new detector technology, however, might overcome the barrier and offer precision tests of $\theta_{23}$. The new detection technology might also lead to a discovery of the CP violating phase in the framework of 3-flavour neutrino oscillations~\cite{Blennow:2015cmn}, which complement the study at T2K and NO$\nu$A to solve the degeneracy problem and exclude CP conserved phase at a relatively high confidence level. In addition, a comprehensive study of the bounds on NSI parameters has been carried out. The bounds on NSI parameters governing the neutrino productions and detections are about one order of magnitude stronger than those related to neutrino propagation in matter, taking the current bounds on $\epsilon^{ud}$ and $\epsilon^{\mu e}$ as an example~\cite{Biggio:2009nt}:
\begin{equation}
|\epsilon^{\mu e}|<
  \left(
  \begin{array}{ccc}
  0.025 & 0.03 & 0.03\\
  0.025 & 0.03 & 0.03\\
  0.025 & 0.03 & 0.03\\
  \end {array}
  \right)\,~ 90\% \textrm{C.L.} \label{equ:bounds1}
\end{equation}
\begin{equation}
|\epsilon^{u d}|<
  \left(
  \begin{array}{ccc}
  0.041 & 0.025 & 0.041\\
  0.026 & 0.078 & 0.013\\
  0.12 & 0.013 & 0.13\\
  \end {array}
  \right)\,~ 90\% \textrm{C.L.} \label{equ:bounds2}
\end{equation}
A special case study was performed at the Daya Bay reactor neutrino experiment to constrain NSI parameters where neutrinos are produced by beta decays and detected by inverse beta decays~\cite{Agarwalla:2014bsa}. The relevant ee sector of $\epsilon^{\mu e}$ got an upper limit of O($10^{-3}$).
With the help of a perturbation theory, neutrino oscillation probabilities in the presence of source/detector and matter NSIs can be found in the reference~\cite{Kopp:2007ne}, which is motivating further study and optimization of new experimental proposals to pin down the current bounds. The first glimpse of NSI effects during neutrino propagation in matter at MOMENT has been shown in the reference~\cite{Bakhti:2016prn}. It has discussed the sensitivity of neutral-current NSIs by means of accelerator neutrino oscillations in matter. However, the impact of source and detector NSIs associated with charged-current interactions has not been discussed.
Within a theoretical model predicting new neutrino interactions, it is natural and fair for neutrinos to carry new charged-current and neutral-current interactions during the production process, the propagation process and the detection process. Furthermore, the current neutrino experiment T2K and NO$\nu$A are based on the superbeam neutrino production where neutrinos come from pion decays. We have to stress that NSIs associated with muon decays are very different from those happening at pion decays if we take the new physics into account. Therefore, it is necessary to bring source/detector NSIs for a complete analysis at MOMENT where neutrinos are produced by muon decays.

In this work, we explore the charged current NSIs effects at MOMENT. We focus on the precision measurement of standard neutrino mixing parameters and constraints of NSI parameters in the presence of non-standard charged-current interactions at the source and detector. The paper is organized as follows: we discuss neutrino oscillation channels at a short and long distance in Section~\ref{sec:oscP}. In Section~\ref{sec:simulation}, we describe our implementations of MOMENT and details in simulation. In Section~\ref{sec:results}, we show the impacts of NSIs on precision measurements of standard neutrino parameters and present the correlations and constraints of NSI parameters within production and detection at MOMENT, and compare the expected results with current bounds. The summary follows in Section~\ref{sec:summary}.

\section{Discussion of neutrino oscillation channels}
\label{sec:oscP}
The formalism of NSI is a general way of studying the impacts of new physics in neutrino oscillations. 
Without dealing with the matter NSIs due to the short baseline, we start with the neutrino production and detection processes involving non-standard interactions. These processes are often related to the charged lepton and called the charged-current-like NSIs. 
The neutrinos at the MOMENT experiment are produced by the muon decay processes $\mu^-\rightarrow e^-+\bar{\nu}_e+\nu_{\mu}$ and $\mu^+\rightarrow e^+ +\nu_e +\bar{\nu}_{\mu}$ and are detected mainly through quasielastic charged-current interactions: $\nu_{\ell}+n\rightarrow p+\ell^-$ and $\bar{\nu}_{\ell}+p\rightarrow n + \ell^+$ (Here $\ell$ denotes e or $\mu$) in the neutrino detector. The CC-NSIs imposed on the production and detection are two different types: the NSIs involved in the muon decay production process are related to charged leptons, while the NSIs involved in the detection process are associated with quarks.
For simplicity, we have restricted the 
operators to $(V-A)(V-A)$ Lorentz structure and neglected NSIs including right handed neutrinos, where the process is helicity suppressed. One may ask why we ignore other potential Lorentz structures, since the most general way of constructing the four-fermion interactions could come with the current of $(V\pm A)(V\pm A)$, $(S\pm P)(S\pm P)$ and $TT$, where $V$ stands for vector couplings, $A$ for axial-vector couplings, $S$ for scalar couplings, $P$ for pseudo-scalar couplings and $T$ for the tensor couplings. There have been several attempts in the literature towards the chirality discussion of NSIs (see e.g.~\cite{Kopp:2007mi,Kopp:2008ds,Kopp:2009zza}). Except $(V-A)(V-A)$, other structures are either helicity suppressed or very small due to their contributions by higher order corrections. Therefore, the most interesting CC-NSIs could be parameterized as the effective four-fermion Lagrangians at the detector: 
\begin{equation}\label{equ:detector_nsi}
\mathcal{L}_{CC-NSI}^{d} = \frac{G_F}{\sqrt{2}} \epsilon_{\alpha\beta}^{d}[\bar{\nu}_{\beta}\gamma^{\mu}(1-\gamma^5)\ell_{\alpha}][\bar{f}\gamma_{\mu}(1-\gamma^5)f']+h.c.
\end{equation}
where the superscript ``d'' represents the interactions at the detector, $G_F$ is the Fermi constant, $f$ ($f'$) represents d (u) quark, $\alpha$ ($\alpha$ = e, $\mu$, $\tau$) is neutrino index and $\beta$ ($\beta$ = e, $\mu$) is lepton index. The NSIs at the detector are parametrized by $\epsilon_{\alpha\beta}^d$ which give the strength of NSIs relative to $G_F$. 
The detector projects the neutrino wave function not only onto the standard weak eigenstates, but onto a combination of them:
\begin{equation}
\langle \nu_\beta^d| = \langle\nu_{\beta}|  +\sum_{\alpha=e,\mu,\tau} \epsilon^d_{\alpha \beta}  \langle \nu_\alpha|
\end{equation}
 In a muon-decay accelerator neutrino experiment, the neutrinos are produced by the muon decay process $\mu^+\rightarrow e^+ +\nu_e+\bar{\nu}_{\mu}$ and the charge conjugated
 process in the Standard Model. The effective lagrangian involving NSIs in production processes can be expressed as:
 \begin{equation}
\mathcal{L}_{CC-NSI}^{s} = \frac{G_F}{\sqrt{2}} \epsilon_{\gamma\delta}^s[\bar{\nu}_{\delta}\gamma^{\mu}(1-\gamma^5)\ell_{\gamma}][\bar{f}\gamma_{\mu}(1-\gamma^5)f^\prime]+h.c.
\end{equation}
where the superscript ``s'' represents the interactions at the source and two fermion fields ``$f$'' 
and ``$f^\prime$'' stand for a neutrino or a charged lepton in order to avoid the confusion of the neutrino flavour index and the index related to the $\gamma$ matrix. For simplicity, we assume that the dominant NSI processes can interfere coherently with standard oscillations. Here we only consider two separate cases (take the $\mu^+$ decay as an example): \\
(1) $\mu^+\to e^+ + \nu_\alpha + \bar{\nu}_\mu$ for NSIs $\epsilon^s_{e \alpha}$ with any flavour ($\alpha=e,\mu,\tau$) by fixing $\bar{\nu}_\mu$, where we assume $f=\mu$, $f^\prime=\nu_{\mu}$, $\ell_{\gamma}=e$.\\
(2) $\mu^+\to e^+ + \nu_e + \bar{\nu}_\alpha$ for NSIs $\epsilon^s_{\mu \alpha}$ with any flavour ($\alpha=e,\mu,\tau$) by fixing $\nu_e$, where we assume $f=e$, $f'=\nu_e$, $\ell_{\gamma}=\mu$.\\
Otherwise, the incoherent process $\mu^+\rightarrow e^+ + \nu_{\alpha}+\bar{\nu}_{\beta}$ ($\alpha,\beta=e,\mu,\tau$) might occur with arbitrary choice of $\nu_{\alpha}$ and $\bar{\nu}_{\beta}$~\cite{Grossman:1995wx, GonzalezGarcia:2001mp, Huber:2002bi, Meloni:2009cg, Blennow:2015nxa}. However, those incoherent contributions to the probabilities are very small since they are suppressed by at least an order of $|\epsilon|^2$ where the SM weak interactions are completely replaced by two NSI vertices. 

Similarly, the neutrino flavour states produced at the source can be written as superpositions of pure flavour eigenstates:
\begin{equation}
|\nu_\delta^s\rangle = |\nu_{\delta}\rangle  +\sum_{\gamma=e,\mu,\tau} \epsilon^s_{\delta\gamma}  |\nu_\gamma\rangle 
\end{equation}
Thus, the oscillation probability is given by:
\begin{equation}
\begin{array}{l}
 P(\nu_\delta^s　\rightarrow　\nu_{\beta}^d) = |\langle  \nu_\beta^d |e^{-i \mathcal{H} L}|\nu_\delta^s  \rangle|^2\\
 ~~~~~~~~~~~~~~~~= |(1+\epsilon^d)_{\eta\beta}(e^{-i \mathcal{H} L})_{\eta \lambda}(1+\epsilon^s)_{\delta\lambda}|^2\\
 ~~~~~~~~~~~~~~~~=|[(1+\epsilon^d)^T e^{-i \mathcal{H} L}(1+\epsilon^s)^T]_{\beta \delta}|^2
\end{array}
\end{equation}
Here the Hamiltonian takes the form of $\mathcal{H}=U$diag$(E+\frac{m_1^2}{2E}, E+\frac{m_2^2}{2E}, E+\frac{m_3^2}{2E})U^{\dagger}$ and U is the PMNS mixing matrix relating the neutrino flavour eigenstates to mass eigenstates $|\nu_{\alpha}\rangle =\sum_{i} U^*_{\alpha i}|\nu_i\rangle$. The $\epsilon^s$ and $\epsilon^d$ are the charged-current NSI matrices for the production and detection, respectively. There are 18 NSI real parameters in total because each complex element $\epsilon_{\alpha \beta}^{s/d}$ consists of the amplitude $|\epsilon_{\alpha\beta}^{s/d}|$ and the phase $\phi_{\alpha\beta}^{s/d}$.

In principle, accelerator neutrinos pass through the earth matter until reaching a detector and matter effects change the oscillation probability. However, MOMENT is a medium baseline experiment and the matter effects are relatively small. Meanwhile, the main topic in the current study is delivered to NSIs happening at the production and detection processes rather than NSIs in matter. Therefore, we will only, for sake of simplicity, display the probabilities perturbatively in vacuum for related appearance and disappearance channels and try to extract useful information for physics performance study. Of course, matter effects are taken into account simultaneously in the complete simulation for physics performance. Since near and far detectors will be used in the simulation later, we will discuss the oscillation channels at a short and far distance separately.

\subsection{Oscillation channels at a near detector}
Here a near detector means detecting neutrinos at a distance of O(100) meters. In the standard oscillation frame without non-standard interactions, ${\nu}_{\mu}$($\bar{\nu}_{\mu}$) and $\nu_e$($\bar{\nu}_e$) can not develop neutrino oscillation patterns in such a short distance and their probabilities are equal to 1. However, NSIs are able to generate \textbf{zero-distance effects} so that the disappearance probabilities are allowed to be larger than 1, equal to 1, or smaller than 1. After dropping the terms $O(\epsilon^2)$, we approximate the probabilities as 
\equ{near_pee_antipee} and \equ{near_pmm_antipmm}:
\begin{equation}\label{equ:near_pee_antipee}
P^{ND}_{\nu_e^s\rightarrow\nu_e^d}(P^{ND}_{\bar{\nu}_e^s\rightarrow\bar{\nu}_e^d}) \approx 1 + 2|\epsilon_{ee}^s|\cos\phi_{ee}^s+2|\epsilon_{ee}^d|\cos\phi_{ee}^d
\end{equation}
\begin{equation}\label{equ:near_pmm_antipmm}
P^{ND}_{\nu_\mu^s\rightarrow\nu_\mu^d}(P^{ND}_{\bar{\nu}_\mu^s\rightarrow \bar{\nu}_\mu^d}) \approx 1 + 2|\epsilon_{\mu\mu}^s|\cos\phi_{\mu\mu}^s+2|\epsilon_{\mu\mu}^d|\cos\phi_{\mu\mu}^d
\end{equation}
It is easy to see that $P(\bar{\nu}_e^s \rightarrow \bar{\nu}_e^d)$/$P(\bar{\nu}_{\mu}^s \rightarrow \bar{\nu}_{\mu}^d)$ deviates from unity with some constant terms in the presence of relevant NSI parameters $\epsilon_{ee}^s$ and $\epsilon_{ee}^d$ ($\epsilon_{\mu\mu}^s$ and $\epsilon_{\mu\mu}^d$). If neutrinos are produced with charged lepton decays and detected by identifying the same charged leptons, the contribution of $\epsilon_{ee}^s$ to the probability is equivalent to $\epsilon_{ee}^d$ and then the sensitivity to these two parameters should be the same at the near detector.

Similarly, the appearance channels $\nu_{\mu}\rightarrow\nu_e$ ($\nu_e\rightarrow\nu_{\mu}$) must remain zero in the standard 3-flavor neutrino scenario. After dropping the $O(\epsilon^3)$ and $O(\epsilon^4)$ terms, the expressions for $\nu_e$ and $\nu_{\mu}$ appearance 
probabilities including CC-NSIs can be written as \equ{near_pme_antipme} and \equ{near_pem_antipem}, respectively:
\begin{equation}\label{equ:near_pme_antipme}
P_{\nu_{\mu}^s \rightarrow \nu_e^d}^{ND}(P_{\bar{\nu}_{\mu}^s \rightarrow \bar{\nu}_e^d}^{ND}) \approx |\epsilon_{\mu e}^s|^2 + |\epsilon_{\mu e}^d|^2 + 2|\epsilon_{\mu e}^s| |\epsilon_{\mu e}^d|\cos(\phi_{\mu e}^s - \phi_{\mu e}^d)
\end{equation}
\begin{equation}\label{equ:near_pem_antipem}
P_{\nu_e^s \rightarrow \nu_{\mu}^d}^{ND}(P_{\bar{\nu}_e^s \rightarrow \bar{\nu}_{\mu}^d}^{ND}) \approx |\epsilon_{e\mu}^s|^2 + |\epsilon_{e\mu}^d|^2 + 2|\epsilon_{e\mu}^s| |\epsilon_{e\mu}^d|\cos(\phi_{e\mu}^s - \phi_{e\mu}^d)
\end{equation}
Here the $\nu_e$($\nu_{\mu}$) appearance probability would depend on $\epsilon_{\mu e}^s$ and $\epsilon_{\mu e}^d$($\epsilon_{e\mu}^s$ and $\epsilon_{e\mu}^d$) after we introduce the NSIs at the neutrino source and detector. The probability of each oscillation channel and its conjugate partner shares the same form at the near detector.  In fact, there are only four effective channels even though eight neutrino oscillation channels can get involved in the MOMENT experiment. It is a discovery of new physics to observe zero-distance effects at near detectors for disappearance or appearance channels. 

\subsection{Oscillation channels at a far detector}

Oscillation patterns get more complicated as soon as we consider channels suitable for the far detector at MOMENT. In the standard framework describing three neutrino mixings, the probability of  $\nu_{\mu}\rightarrow\nu_e$ channel is calculated by a simple change of sign of the $\sin\delta$ term in a T-reversed channel of $\nu_e\rightarrow\nu_{\mu}$. Due to CC-NSIs, $\nu_{\mu}\rightarrow\nu_e$ and $\nu_e\rightarrow\nu_{\mu}$ probabilities are not so obvious any more. We perturbatively derive the explicit expressions of their probabilities in vaccum as given in \equ{far_pme_invacuum} and \equ{far_pem_invacuum}, considering $\alpha=\frac{\Delta m_{21}^2}{\Delta m_{31}^2}\approx0.03$, $s_{13}=\sin\theta_{13}\approx0.15$ and NSI parameters as small numbers. In order to clearly show the impacts of NSIs, we can split the $P_{\nu_\mu \rightarrow \nu_e}^{FD}$ ($P_{\nu_e\rightarrow \nu_{\mu}}^{FD}$) into a sum of three terms: the standard oscillation
term  $P_{\nu_\mu \rightarrow \nu_e}^{SM}$ ($P_{\nu_e \rightarrow \nu_{\mu}}^{SM}$), the dominant order of $\mathcal{O}$($\epsilon s_{13}$) NSI oscillatory term $P_{\nu_\mu \rightarrow \nu_e}^{NSI(\epsilon s_{13})}$ ($P_{\nu_e \rightarrow \nu_{\mu}}^{NSI(\epsilon s_{13})}$) and the sub-dominant order of $\mathcal{O}$($\alpha\epsilon$) NSI oscillatory term $P_{\nu_\mu \rightarrow \nu_e}^{NSI(\alpha\epsilon)}$ ($P_{\nu_e \rightarrow \nu_{\mu}}^{NSI(\alpha\epsilon)}$).

For an oscillation channel of $\nu_{\mu}\rightarrow\nu_e$, the probability can be written as:
\begin{eqnarray}
P_{\nu_\mu \rightarrow \nu_e}^{FD} && = P_{\nu_\mu \rightarrow \nu_e}^{SM}+ P_{\nu_\mu \rightarrow \nu_e}^{NSI(\epsilon s_{13})}+P_{\nu_\mu \rightarrow \nu_e}^{NSI(\alpha\epsilon)}+\mathcal{O}(\alpha^3)+\mathcal{O}(\alpha^2s_{13})+\mathcal{O}(\alpha s_{13}^2)+\mathcal{O}(s_{13}^3)+\mathcal{O}(\epsilon\alpha^2)\nonumber\\
&&+\mathcal{O}(\epsilon s_{13}^2)+\mathcal{O}(\epsilon^2),
\label{equ:far_pme_invacuum}
\end{eqnarray} 
with
\begin{eqnarray}
P_{\nu_\mu \rightarrow \nu_e}^{SM} &&\approx s_{2\times13}^2 s_{23}^2 \sin^2\Delta_{31}+\alpha\Delta_{31}s_{2\times12}s_{2\times23}s_{13}[\sin(2\Delta_{31})\cos\delta - 2\sin\delta\sin^2\Delta_{31}]+\alpha^2\Delta_{31}^2c_{23}^2s_{2\times12}^2,
\label{equ:far_pme_1}
\end{eqnarray} 
\begin{eqnarray}
P_{\nu_\mu \rightarrow \nu_e}^{NSI(\epsilon s_{13})} &&\approx-2s_{2\times13}s_{23}[|\epsilon_{\mu e}^s|\cos(\delta+\phi_{\mu e}^s)+c_{2\times23}|\epsilon_{\mu e}^d|\cos(\delta+\phi_{\mu e}^d)-s_{2\times23}|\epsilon_{\tau e}^d|\cos(\delta+\phi_{\tau e}^d)] \sin^2\Delta_{31}\nonumber\\
&&-s_{2\times13} s_{23}[|\epsilon_{\mu e}^s|\sin(\delta+\phi_{\mu e}^s)+|\epsilon^d_{\mu e}|\sin(\delta+\phi^d_{\mu e})]\sin(2\Delta_{31})\label{equ:far_pme_2},
\end{eqnarray} 
\begin{eqnarray}
P_{\nu_\mu \rightarrow \nu_e}^{NSI(\alpha\epsilon)} &&\approx+2\alpha \Delta_{31} |\epsilon_{\mu e}^d| s_{2\times12} c_{13} c_{23} s_{23}^2 \cos\phi_{\mu e}^d \sin(2\Delta_{31})+2\alpha \Delta_{31} |\epsilon_{\tau e}^d| c_{13} c_{23}^2 s_{2\times12} s_{23} \cos\phi_{\tau e}^d \sin(2\Delta_{31})\nonumber\\
&&-2\alpha \Delta_{31} |\epsilon_{\mu e}^d| s_{2\times12} c_{13} c_{23} \sin\phi_{\mu e}^d (1-2 s_{23}^2 \sin^2\Delta_{31})+4\alpha \Delta_{31} |\epsilon_{\tau e}^d| c_{13} c_{23}^2 s_{2\times12} s_{23} \sin\phi_{\tau e}^d \sin^2\Delta_{31}\nonumber\\
&&-2\alpha \Delta_{31} |\epsilon_{\mu e}^s| c_{13} s_{2\times12} c_{23} \sin\phi_{\mu e}^s.
\label{equ:far_pme_3}
\end{eqnarray}
In a similar way, the probability of $\nu_e\rightarrow\nu_{\mu}$ can be expressed as:
\begin{eqnarray}
P_{\nu_e \rightarrow \nu_{\mu}}^{FD} && = P_{\nu_e \rightarrow \nu_{\mu}}^{SM}+ P_{\nu_e \rightarrow \nu_{\mu}}^{NSI(\epsilon s_{13})}+P_{\nu_e \rightarrow \nu_{\mu}}^{NSI(\alpha\epsilon)}+\mathcal{O}(\alpha^3)+\mathcal{O}(\alpha^2s_{13})+\mathcal{O}(\alpha s_{13}^2)+\mathcal{O}(s_{13}^3)+\mathcal{O}(\epsilon\alpha^2)\nonumber\\
&&+\mathcal{O}(\epsilon s_{13}^2)+\mathcal{O}(\epsilon^2),
\label{equ:far_pem_invacuum}
\end{eqnarray} 
with
\begin{eqnarray}
P_{\nu_e \rightarrow \nu_{\mu}}^{SM} &&\approx s_{2\times13}^2 s_{23}^2 \sin^2\Delta_{31}+\alpha\Delta_{31}s_{2\times12}s_{2\times23}s_{13}[\sin(2\Delta_{31})\cos\delta+2\sin\delta\sin^2\Delta_{31}]+\alpha^2\Delta_{31}^2c_{23}^2s_{2\times12}^2,
\label{equ:far_pem_1}
\end{eqnarray} 
\begin{eqnarray}
P_{\nu_e \rightarrow \nu_{\mu}}^{NSI(\epsilon s_{13})} &&\approx-2s_{2\times13}s_{23}[|\epsilon_{e\mu}^d|\cos(\delta-\phi_{e\mu}^d)+c_{2\times23}|\epsilon_{e\mu}^s|\cos(\delta-\phi_{e\mu}^s)-s_{2\times23}|\epsilon_{e\tau}^s|\cos(\delta-\phi_{e\tau}^s)]\sin^2\Delta_{31}\nonumber\\
&&+s_{2\times13}s_{23}[|\epsilon_{e\mu}^s|\sin(\delta-\phi_{e\mu}^s)+|\epsilon_{e\mu}^d|\sin(\delta-\phi_{e\mu}^d)]\sin(2\Delta_{31}),
\label{equ:far_pem_2}
\end{eqnarray} 
\begin{eqnarray}
P_{\nu_e \rightarrow \nu_{\mu}}^{NSI(\alpha\epsilon)} &&\approx +2\alpha \Delta_{31} |\epsilon_{e\mu}^s| s_{2\times12} c_{13} c_{23} s_{23}^2 \cos\phi_{e\mu}^s \sin(2\Delta_{31})+2\alpha \Delta_{31} |\epsilon_{e\tau}^s| c_{13} c_{23}^2 s_{2\times12} s_{23} \cos\phi_{e\tau}^s \sin(2\Delta_{31})\nonumber\\
&&-2\alpha \Delta_{31} |\epsilon_{e\mu}^s| s_{2\times12} c_{13} c_{23} \sin\phi_{e\mu}^s (1-2s_{23}^2 \sin^2\Delta_{31})+4\alpha \Delta_{31} |\epsilon_{e\tau}^s| c_{13} c_{23}^2 s_{2\times12} s_{23} \sin\phi_{e\tau}^s \sin^2\Delta_{31}\nonumber\\
&&-2\alpha \Delta_{31} |\epsilon_{e\mu}^d| c_{13} s_{2\times12} c_{23} \sin(\phi_{e\mu}^d).
\label{equ:far_pem_3}
\end{eqnarray} 
Here $s_{ij}=\sin\theta_{ij}, c_{ij}=\cos\theta_{ij}, s_{2\times ij}=\sin2\theta_{ij}, c_{2\times ij}=\cos2\theta_{ij}, \Delta_{31}=\frac{\Delta m_{31}^2 L}{4 E},
\alpha=\frac{\Delta m_{21}^2}{\Delta m_{31}^2}$ and $\epsilon_{\alpha\beta}^{s/d}=|\epsilon_{\alpha\beta}^{s/d}| e^{i\phi_{\alpha\beta}^{s/d}}$. 
\begin{figure}[!t]
\includegraphics[scale=0.65]{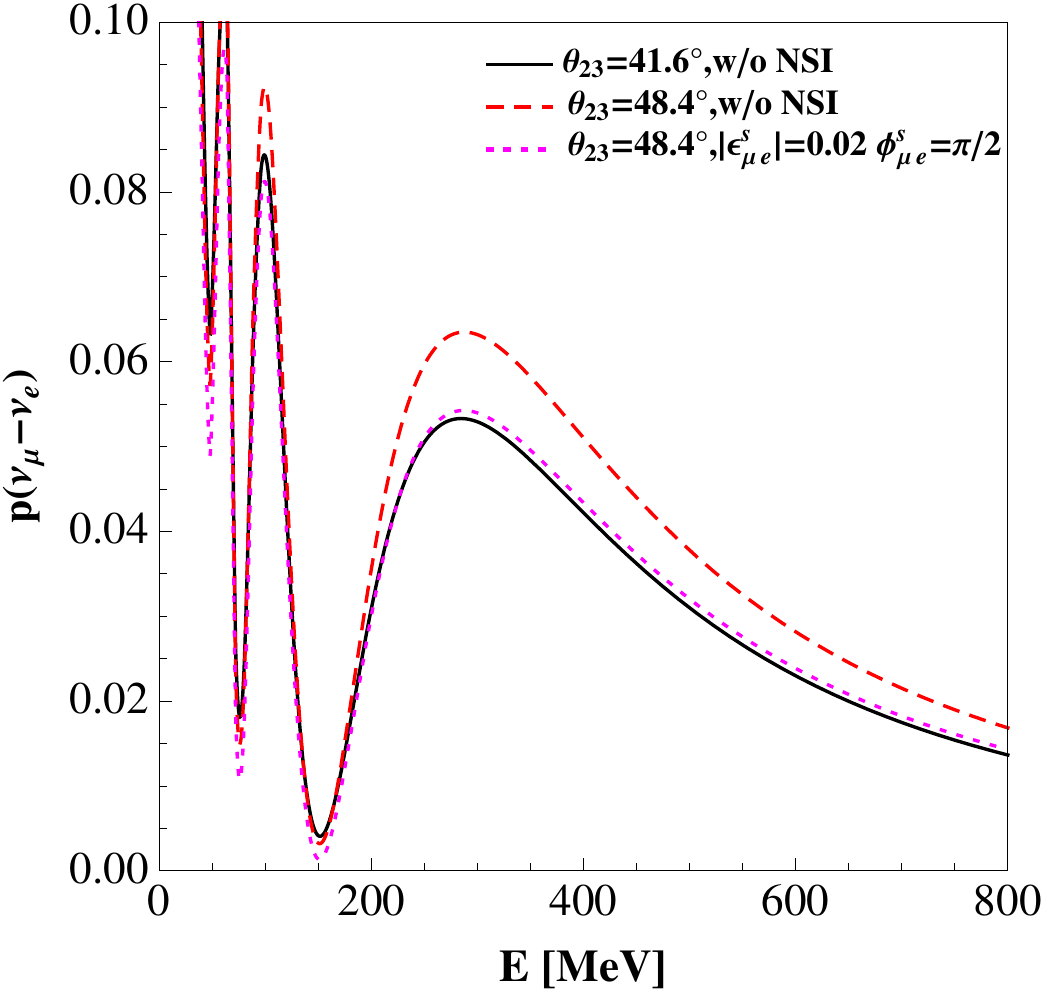}
\caption{The oscillation probability of $P(\nu_{\mu}\rightarrow\nu_e)$ as a function of neutrino energy. The parameter settings for different lines are interpreted in the legend. Here we assume the normal mass hierarchy and take the following inputs: $\Delta m_{21}^2=7.5\times10^{-5}$ eV$^2$, $\theta_{12}=33.56^\circ$, $\Delta m_{31}^2=2.524\times10^{-3}$ eV$^2$, $\theta_{13}=8.46^\circ$ and $\delta=270^\circ$.}
\label{fig:octant_prob}
\end{figure}
Our derived oscillation probabilities are consistent with the reference~\cite{Kopp:2007ne}. We can immediately read the impacts of NSI parameters on appearance channels in order:
\begin{itemize}
 \item {It is clear that $\nu_{\mu}\rightarrow\nu_e$ ($\nu_e\rightarrow\nu_{\mu}$) is affected by the dominant NSI parameters $\epsilon_{\mu e}^s$, $\epsilon_{\mu e}^d$ and $\epsilon_{\tau e}^d$ ($\epsilon_{e\mu}^s$, $\epsilon_{e\tau}^s$ and $\epsilon_{e\mu}^d$). When the dominant term of $\mathcal{O}$($s_{13}\epsilon$) is considered, turning on $\epsilon_{\mu e}^s$
and $\epsilon_{\tau e}^d$ ($\epsilon_{e\mu}^d$ and $\epsilon_{e\tau}^s$) is approximately equivalent to enlarging or compressing the amplitude of $\sin^2\Delta_{31}$. Meanwhile, the term of $\epsilon_{\mu e}^d$ ($\epsilon_{e\mu}^s$) will change the maximal position of its oscillation probability.}
 \item{When $\theta_{23}$ approaches 45 degrees, $s_{2\times23}\approx1$ and $c_{2\times23}\approx0$. Thus $\epsilon_{\mu e}^d$ ($\epsilon_{e\mu}^s$) would lose the effects on the term of $\sin^2\Delta_{31}$ related to $\mathcal{O}(\epsilon s_{13})$ for the $\nu_{\mu}\rightarrow\nu_e$ ($\nu_{e}\rightarrow\nu_{\mu}$) channel. }
 \item{Since these NSI parameters are entangled with standard mixing parameters, NSIs would interfere with precision measurements of the standard CP-violating phase $\delta$ and $\theta_{23}$ which is manifested by related terms in \equ{far_pme_invacuum} and \equ{far_pem_invacuum}. These two channels are sensitive to the octant of $\theta_{23}$ beacuse the leading order of \equ{far_pme_invacuum} and \equ{far_pem_invacuum} depend on $\sin^2\theta_{23}$. The presence of NSIs may induce a wrong determination of $\theta_{23}$. Fig.~\ref{fig:octant_prob} shows a specific situation of the degeneracy caused by NSIs at the probability level. The  continuous line and the dashed line show the case of $\theta_{23}=41.6^{\circ}$ and $\theta_{23}=48.4^{\circ}$ without NSI, respectively. We can see a clear separation between two scenarios. However, introducing NSIs can shift the amplitude of the probability and fake the contribution of $\theta_{23}$. For instance, the dotted line shows the special case with $|\epsilon_{\mu e}^s|=0.02$ and $\phi_{\mu e}^s=\pi/2$. It is clear that the dotted oscillation curve almost coincides with the continuous curve. This indicates that we may get the wrong measurement of $\theta_{23}$, if there is indeed CC-NSIs but we ignore them by only fitting the data within the standard neutrino oscillation framework.}
 \item{An assumption of real NSI parameters with all NSI phases $\phi=0$ could leads to a substantial simplification in each channel, since those terms proportional to $\sin\phi_{\alpha\beta}^{s/d}$ would vanish. One can observe that the sensitivity to $\epsilon_{\tau e}^d$ ($\epsilon_{e\tau}^s$) from the oscillation channel $\nu_{\mu}\rightarrow\nu_e$ ($\nu_e\rightarrow\nu_{\mu}$) would be tiny for $\delta = 3\pi/2$, $\pi/2$. Moreover, we can obtain a linear correlation of $\epsilon_{\mu e}^s$ and $\epsilon_{\mu e}^d$ (or $\epsilon_{e\mu}^s$ and $\epsilon_{e\mu}^d$). In the case of $\delta=0/\pi$, $\sin\delta$ term will vanish and we can get the linear correlation of $\epsilon_{\mu e}^s$, $\epsilon_{\mu e}^d$ and $\epsilon_{\tau e}^d$ ($\epsilon_{e\mu}^s$, $\epsilon_{e\mu}^d$ and $\epsilon_{e\tau}^s$). Futhermore, if $\theta_{23}$ is close to 45 degrees, the contributions from $\epsilon_{\mu e}^d$ ($\epsilon_{e\mu}^s$) tend to disappear. In brief, the sensitivities to the related CC-NSI parameters at a far detector vary with the standard parameters $\theta_{23}$ and $\delta$.}
\end{itemize}   

Similar to the discussion above, $\nu_{\mu}$ and $\nu_e$ disappearance probabilities are given as \equ{far_pmm_antipmm} and \equ{far_pee_antipee}:
\begin{eqnarray}
   P_{\nu_{\mu}^s\rightarrow\nu_{\mu}^d}^{FD}&& \approx 1
-s_{2\times23}^2\sin^2\Delta_{31}
+2|\epsilon_{\mu\mu}^s|\cos\phi_{\mu\mu}^s+2|\epsilon_{\mu\mu}^d|\cos\phi_{\mu\mu}^d +s_{2\times23}(|\epsilon_{\mu\tau}^s|\sin\phi_{\mu\tau}^s+|\epsilon_{\tau\mu}^d|\sin\phi_{\tau\mu}^d)\sin(2\Delta_{31})\nonumber\\
&&-2s_{2\times23}^2(|\epsilon_{\mu\mu}^s|\cos\phi_{\mu\mu}^s+|\epsilon_{\mu\mu}^d|\cos\phi_{\mu\mu}^d)\sin^2\Delta_{31}
-2c_{2\times23}s_{2\times23}(|\epsilon_{\mu\tau}^s|\cos\phi_{\mu\tau}^s+|\epsilon_{\tau\mu}^d|\cos_{\tau \mu}^d)\sin^2\Delta_{31}
\label{equ:far_pmm_antipmm}
\end{eqnarray}
\begin{eqnarray}
  P_{\nu_e^s\rightarrow\nu_e^d}^{FD}&&\approx 1-4s_{13}^2\sin^2\Delta_{31}
 +2|\epsilon_{ee}^s|\cos\phi_{ee}^s+2|\epsilon_{ee}^d|\cos\phi_{ee}^d \nonumber\\
&& -2s_{13}s_{23}[|\epsilon_{e\mu}^s|\sin(\delta-\phi_{e\mu}^s)-|\epsilon_{\mu e}^d|\sin(\delta+\phi_{\mu e}^d)]\sin(2\Delta_{31})\nonumber\\
 &&-2s_{13}c_{23}[|\epsilon_{e\tau}^s|\sin(\delta-\phi_{e\tau}^s)-|\epsilon_{\tau e}^d|\sin(\delta+\phi_{\tau e}^d)]\sin(2\Delta_{31})\nonumber\\
 && -4s_{13}s_{23}[|\epsilon_{e\mu}^s|\cos(\delta-\phi_{e\mu}^s)+|\epsilon_{\mu e}^d|\cos(\delta+\phi_{\mu e}^d)]\sin^2\Delta_{31}\nonumber\\
 &&-4s_{13}c_{23}[|\epsilon_{e\tau}^s|\cos(\delta-\phi_{e\tau}^s)+|\epsilon_{\tau e}^d|\cos(\delta+\phi_{\tau e}^d)]\sin^2\Delta_{31}
 \label{equ:far_pee_antipee}
\end{eqnarray}
Apart from standard neutrino oscillations, major contributions come from the terms proportional to $\epsilon_{\mu\mu}^{s/d}$ ($\epsilon_{ee}^{s/d}$) rather than other NSIs in the $\nu_{\mu}$ ($\nu_e$) disappearance channel. We expect better constraints on $\epsilon_{\mu\mu}^{s/d}$ ($\epsilon_{ee}^{s/d}$). 
The $\nu_{\mu}\rightarrow\nu_{\mu}$ ($\bar{\nu}_{\mu}\rightarrow\bar{\nu}_{\mu}$) is an important channel to measure $\theta_{23}$, which is expected to judge whether $\theta_{23}$ is maximal or not. The channel of $\nu_e\rightarrow\nu_e$ ($\bar{\nu}_e\rightarrow\bar{\nu}_e)$ is good at precision measurements of $\theta_{13}$. Without NSIs, $\nu_e$ ($\bar{\nu}_e$) disappearance channel has no dependence of the standard CP phase. After introducing NSI parameters, however, even the standard CP-violating phase would appear in the $\nu_e$ ($\bar{\nu}_e$) disappearance probability. Under the assumption of $\phi=0$, we will have a rather simplified correlation of standard neutrino oscillation and NSI parameters. A few comments for this special case are given as follows:
\begin{itemize}
 \item{The oscillation channel $\nu_{\mu}\rightarrow\nu_{\mu}$ ($\bar{\nu}_{\mu}\rightarrow\bar{\nu}_{\mu}$) is affected by $|\epsilon_{\mu\mu}^{s/d}|$, $|\epsilon_{\mu\tau}^s|$ and $|\epsilon_{\tau\mu}^d|$. Apart from standard neutrino oscillation terms, the major contributions could be associated with the constant terms of $|\epsilon_{\mu\mu}^{s/d}|$. When $\theta_{23}$ is getting closer to 45 degrees, the coefficient of $|\epsilon_{\mu\mu}^{s/d}|$ should be much larger than the coefficient of $|\epsilon_{\mu\tau}^s|$ ($|\epsilon_{\tau\mu}^d|$). Therefore, this channel is more sensitive to $|\epsilon_{\mu\mu}^{s/d}|$ rather than $|\epsilon_{\mu\tau}^s|$ and $|\epsilon_{\tau\mu}^d|$.
 }
 \item {Parameters $\epsilon_{ee}^{s/d}$, $\epsilon_{e\mu}^s$, $\epsilon_{\mu e}^d$, $\epsilon_{e\tau}^s$ and $\epsilon_{\tau e}^d$ affect the $\nu_e$ ($\bar{\nu}_e$) disappearance probability.
$\epsilon_{ee}^{s/d}$ have the most important impacts on the channel. The sensitivities to $\epsilon_{e\mu}^s$, $\epsilon_{\mu e}^d$, $\epsilon_{e\tau}^s$ and $\epsilon_{\tau e}^d$ 
are smaller and depend on $\delta$. On the contrary, these four parameters can be well constrained in appearance channels.}
 \end{itemize}
\begin{figure}[!t]
\includegraphics[scale=0.65]{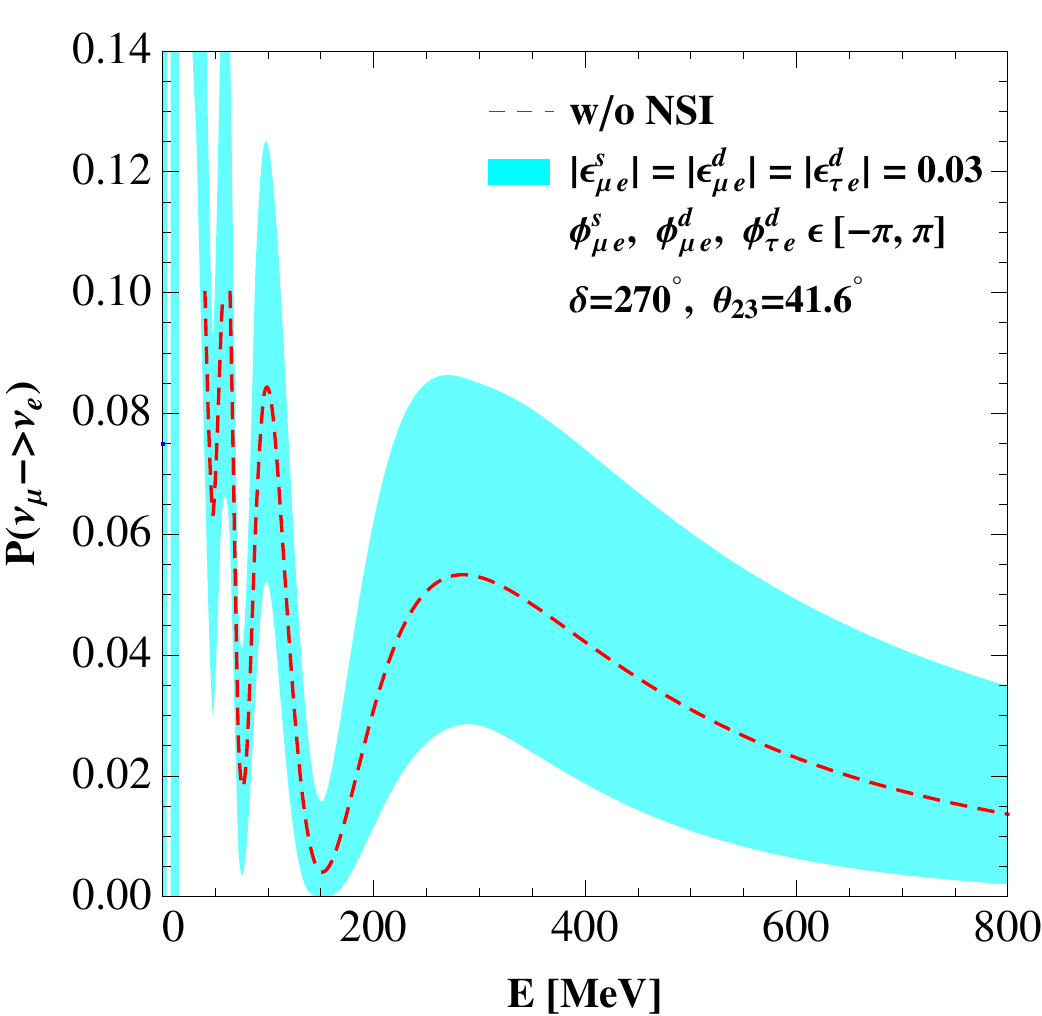}
\includegraphics[scale=0.65]{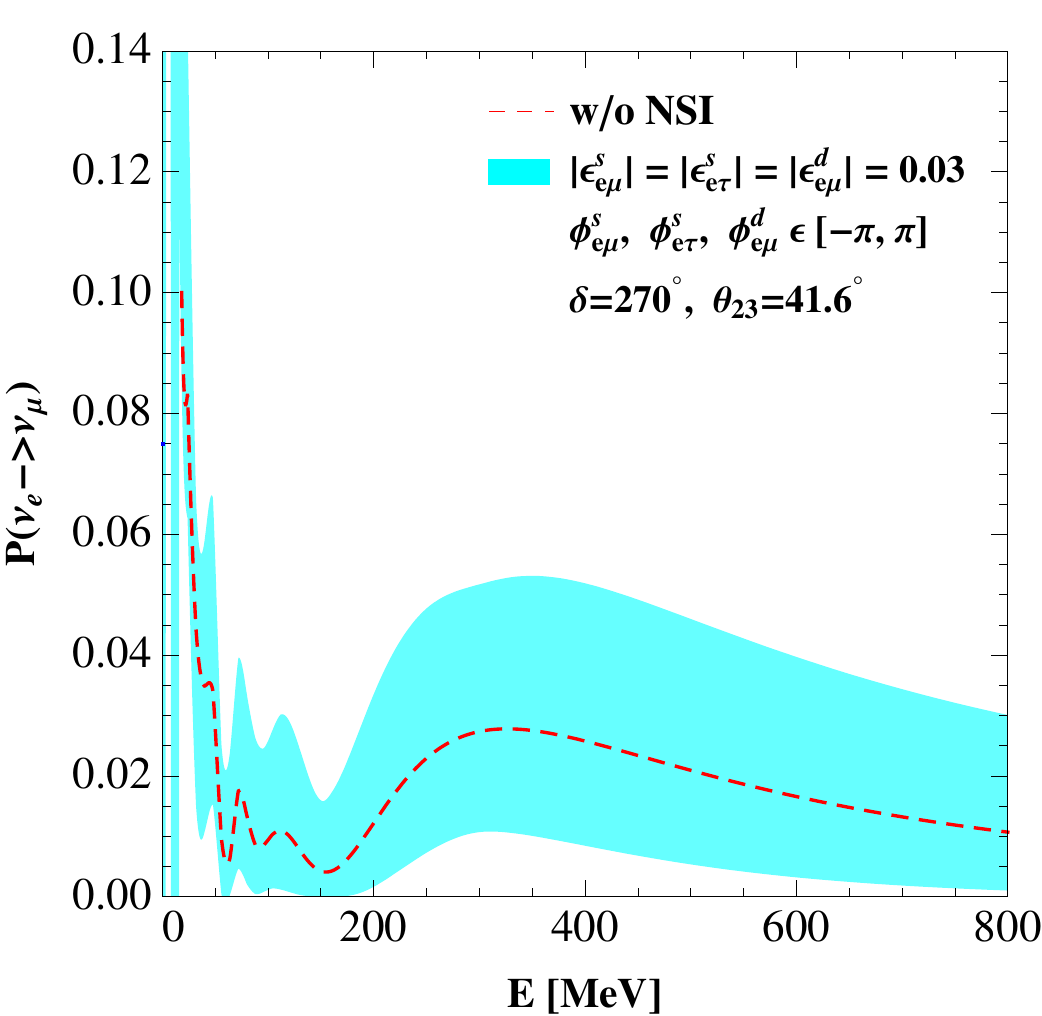}
\caption{The oscillation probabilities of $P(\nu_{\mu}\rightarrow\nu_e)$ and $P(\nu_e\rightarrow\nu_{\mu})$ with/without CC-NSIs at a far detector given the baseline of 150 km. In both panels, the shaded region in cyan is due to the variation of relevant NSI parameters as is shown in the legend. The dashed red line represents the case without NSI parameters. Here we assume the normal mass hierarchy and take the following inputs: $\Delta m_{21}^2=7.5\times10^{-5}$ eV$^2$, $\theta_{12}=33.56^\circ$, $\Delta m_{31}^2=2.524\times10^{-3}$ eV$^2$ and $\theta_{13}=8.46^\circ$.}
\label{fig:probband}
\end{figure}
In Fig.~\ref{fig:probband}, we make a comparison of the probabilities $P(\nu_{\mu}\rightarrow\nu_e)$ and $P(\nu_e\rightarrow\nu_{\mu})$ with/without NSIs. We turn on non-vanishing parameters of $|\epsilon_{\mu e}^s|=|\epsilon_{\mu e}^d|=|\epsilon_{\tau e}^d|=0.03$ for an illustration. The amplitude of oscillation pattern is shifted significantly by these NSI impacts as we point out from the perturbative approximation. We display the probability curve of standard three-flavour neutrino oscillation and the probability bands originating from the variations of NSI parameters. The shaded region in Fig.~\ref{fig:probband} shows changes of oscillation probabilities when the standard CP-violating phase is $3\pi/2$ and the NSI phases vary in $[-\pi, \pi]$. It is easy to see the difference between $P(\nu_{\mu}\rightarrow\nu_e)$ and $P(\nu_e\rightarrow\nu_{\mu})$ without NSI effects, where $P(\nu_{\mu}\rightarrow\nu_e)-P(\nu_e\rightarrow\nu_{\mu})$ becomes negative for $\delta\in(0, \pi)$ and $P(\nu_{\mu}\rightarrow\nu_e)$ is larger than $P(\nu_e\rightarrow\nu_{\mu})$ for $\delta\in(\pi,2\pi)$; with $\delta=0/\pi$, there is no difference between $P(\nu_{\mu}\rightarrow\nu_e)$ and $P(\nu_e\rightarrow\nu_{\mu})$. Once we turn on NSI parameters, however, the relationship between the probabilities of these two channels will be changed significantly as can be seen from the shaded regions. After we introduce the characteristics of MOMENT and simulation details in the next section, we will be further convinced by an analysis of event rates as shown in~\figu{events}. 

\section{Characteristics of MOMENT and the details of simulation}
\label{sec:simulation}

Continuous high-energy proton beam at 1.5 GeV with a power of 15 MW will be produced in the accelerator facility. A choice of target stations with fluidised tungsten is still under optimization to generate charged mesons most of which are pions and kaons. We can expect $1.1\times10^{24}$ Proton On Target (POT) per year. A magnetic solenoid will be deployed to make the pion beam focused and selected. The curvature of the solenoid helps with selecting muons from pion decays, followed by a straight tunnel to prepare neutrinos from muon decays. The neutrino fluxes are provided by the accelerator working group in MOMENT for our physics performance study~\cite{Nikos:2017}. 
We intend to extract more information from eight oscillation channels using the muon-decay neutrino beams in the simulation study: $\nu_e\rightarrow \nu_e$, $\nu_e\rightarrow \nu_{\mu}$, $\nu_{\mu} \rightarrow \nu_e$, $\nu_{\mu} \rightarrow \nu_{\mu}$ and their conjugate partners. Since we have to conduct flavour and charge identifications to distinguish secondary particles, we consider the new technology using Gd-doping water to separate both Cherenkov and coincident signals from capture of thermal neutrons. Muon taggings can be efficiently obtained by daughter electrons together with pulse shape discrimination of waveforms. We follow the detector description from a sophisticated study in the CERN-MEMPHYS project~\cite{Campagne:2006yx} and update the related new technology with regard to Gd-doping water. The major backgrounds for MOMENT come from atmospheric neutrinos. We believe that they can be suppressed by the beam direction and proper modelling background spectra within the beam-off period, which is to be extensively studied in detector simulations in future study. 

\begin{table}[!t]
\centering
\begin{tabular}{|c|c|c|c|c|}
\hline
Parameters & Best-fit values & Prior uncertainties \\
\hline
\tabincell{c}{$\theta_{12}/^{\circ}$}& 33.56 & $2.3\%$\\
\hline
\tabincell{c}{$\theta_{13}/^{\circ}$}& 8.46  &$1.8\%$ \\
\hline
\tabincell{c}{$\theta_{23}/^{\circ}$}&  41.6 &$5.8\%$ \\
\hline
\tabincell{c}{$\Delta m_{21}^2$/eV$^2$}& 7.5$\times10^{-5}$ &$2.4\%$\\
\hline
\tabincell{c}{$\Delta m_{31}^2$/eV$^2$}& 2.524$\times10^{-3}$ &$1.6\%$\\
\hline
\end{tabular}
\caption{The best-fit values of standard parameters and their prior uncertainties adopted in the numerical simulations~\cite{Esteban:2016qun}.}
\label{tab:nu-fitvalues}
\end{table}
\begin{figure}[!t]
\includegraphics[scale=0.65]{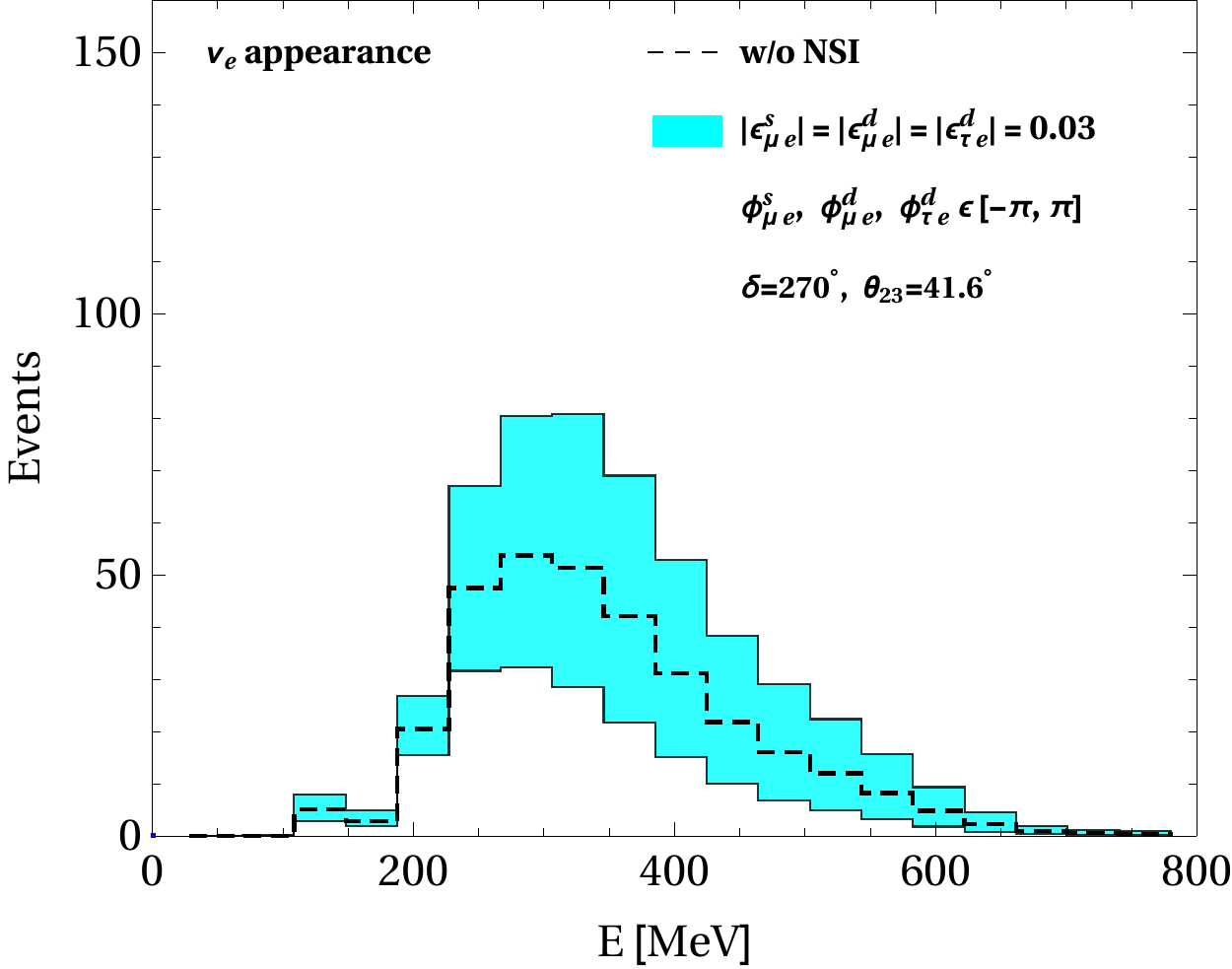}
\includegraphics[scale=0.65]{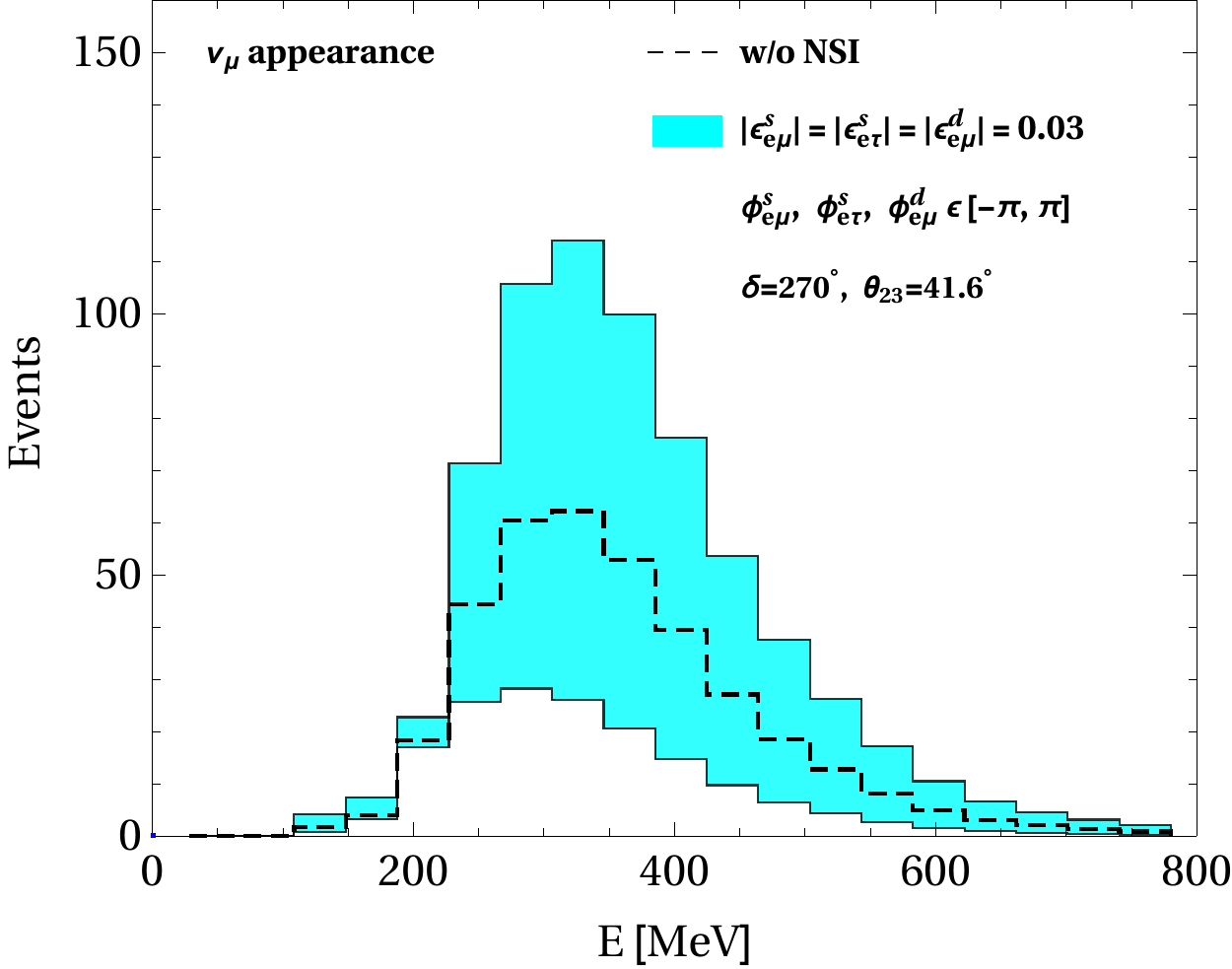}
\caption{The event rates of $\nu_{\mu}\rightarrow\nu_e$ and $\nu_e\rightarrow\nu_{\mu}$ channels versus the neutrino energy. Variations of the relevant NSI parameters give rise to the cyan bands. The dashed black lines represent the case without NSIs.}
\label{fig:events}
\end{figure}
Charged-current interactions are used to identify neutrino signals: 
$$
\nu_e + n \rightarrow p + e^-~~~~~~~~~~\bar{\nu}_{\mu} + p \rightarrow n + \mu^+
$$
$$
\bar{\nu}_e + p \rightarrow n + e^+ ~~~~~~~~~~\nu_{\mu} + n \rightarrow p + \mu^- 
$$
A few remarks for signals and backgrounds are given as follows:
\begin{itemize}
 \item {Gd doping into pure water could be used to discriminate electron neutrinos and antineutrinos by whether there is a capture of the scattered thermal neutron or not. Neutron capture on Gd emits the 8 MeV gamma rays. The $\bar{\nu}_e$ ($\bar{\nu}_{\mu}$) signal is reconstructed by tagging the neutron in coincidence with the positron to suppress most of backgrounds associated with single events. While water Cherenkov detection is not significantly changed, $\nu_e$ signals come from the Cherenkov ring created by $\nu_e$ elastic scattering with electrons.}
 \item {In a Water Cherenkov detector, electron and muon-flavour neutrinos are well separated by event reconstructions, where the former type creates electron showers and the latter type leads to muon tracks. Sometimes, low-energy muons decay and cause flavour misidentifications. Here we ignore the flavour misidentification in the simulation.}
 \item {The imperfection of detectors leads to misidentifications of charge-current interactions for neutrino signals. Here we suppose their effects are negligible.}
 \item {It is also possible for neutral-current interactions with accidental single events to be identified as the coincident signal. A neutron knocks a nucleus off of an oxygen, resulting in excited states and photons from de-excitations will mimick $\bar{\nu}_e$ signals oscillated from accelerator neutrino beams. We expect them to be rather small and assign an extremely small background over signal ratio in our simulation.}
\end{itemize}
Table~\ref{tab:glbtable} lists the simulation details about the neutrino detector. A baseline of 150 km is assigned in the current proposal based on the neutrino beam energy range~\cite{Cao:2014bea}. We assume a near detector with a fiducial mass of 100 t and a far detector with a fiducial mass of 500 kton. The running time is 5 years for each polarity. In the massive water cherenkov detector, we follow electron and muon selection efficiencies given in Ref.~\cite{Agostino:2012fd,Huber:2008yx}. With regard to the normalization error on signals and the normalization error on backgrounds, we assume they are at the level of 5\%. As for the atmospheric backgrounds, they could be suppressed via sending the neutrino beam in short bunches with a suppression factor of $2.2\times10^{-3}$~\cite{FernandezMartinez:2009hb}. 
The cross section for quasi-elastic interactions is taken from the reference~\cite{Paschos:2001np}. 

The values of the standard neutrino oscillation parameters are taken from the latest nu-fit results~\cite{Esteban:2016qun}. Table~\ref{tab:nu-fitvalues} shows the central values and their uncertainties in the present work. Unless otherwise mentioned, we expect a determination of mass hierarchy without NSIs before running MOMENT and assume the normal mass hierarchy in our simulation without a loss of generality, i.e. $\Delta m_{31}^2>0$. A list of assumptions for near and far detectors are given in Table~\ref{tab:glbtable}. We present numerical results by simulating the neutrino oscillation signals and backgrounds using GLoBES~\cite{Huber:2004ka,Huber:2007ji}. Similar to the probability-level analysis given in~\figu{probband}, we present event rates of $\nu_{\mu}\rightarrow\nu_e$ and $\nu_e\rightarrow\nu_{\mu}$ channels versus the neutrino energy in~\figu{events}. The event spectra are shifted significantly after we consider CC-NSI effects. The shaded region highlights the large variation from the new CP phases caused by CC-NSIs even if we fix their strength of NSI couplings. It is then straightforward to discuss physics performance for the MOMENT experiment with simulated event spectra. We calculate the event rates by defining the true values (central values) for standard oscillation parameters and fit with/without NSI impacts to extract useful information. We compute the $\chi^2$ using the following approach:
\begin{eqnarray}
\chi^2=[\sum_j^{channel}\sum_i^{bin}\frac{|N_{ij}(\boldsymbol\rho_{true},\boldsymbol\epsilon_{true})-N_{ij}(\boldsymbol\rho_{test},\boldsymbol\epsilon_{test},\boldsymbol{s})|^2}{N_{ij}(\boldsymbol\rho_{true},\boldsymbol\epsilon_{true})}+\sum_{\alpha}\frac{(\rho_{\alpha}-\rho_{\alpha}^{true})^2}{\sigma^2_{\rho_{\alpha}}}+\sum_{\beta}\frac{(s_{\beta}-s_{\beta}^{true})^2}{\sigma^2_{s_{\beta}}}]_{min},
\label{equ:chi_square}
\end{eqnarray}
where the index j denotes the channel number and i denotes the bin number, $\boldsymbol\rho_{true}$ and $\boldsymbol\rho_{test}$ represent the standard vectors of true and test values, respectively, $\boldsymbol{s}$ is the vector of systematics related to the neutrino beam and detector,  $\boldsymbol\epsilon_{true}$ and $\boldsymbol\epsilon_{test}$ are the non-standard vectors of true values and test values, and $N_{ij}$ is the expected event for the j-th channel and i-th bin. The second term corresponds to the contribution to $\chi^2$ from the external inputs which are based on results from previous experiments. $\sigma_{\rho_{\alpha}}$ is the external error (input error) in GLoBES imposed on the central values. Similarly, the third term represents the treatment of systematic errors implemented on the $\chi^2$.

\begin{table}
\centering
\begin{tabular}{|c|c|c|c|c|}
\hline
\tabincell{c}{Fiducial mass (ND/FD)} &\tabincell{c}{Gd-doping Water cherenkov \\(100 t/500 kton)}     \\
\hline
Baseline (ND/FD) & 500 m/150 km \\
\hline
\tabincell{c}{Channels} & \tabincell{c}{$\nu_e(\bar{\nu}_e)\rightarrow\nu_e(\bar{\nu}_e)$, $\nu_{\mu}(\bar{\nu}_{\mu})\rightarrow\nu_{\mu}(\bar{\nu}_{\mu})$\\$\nu_e(\bar{\nu}_e)\rightarrow\nu_{\mu}(\bar{\nu}_{\mu})$, $\nu_{\mu}(\bar{\nu}_{\mu})\rightarrow\nu_e(\bar{\nu}_{e})$}\\
\hline
Energy resolution &$8.5\%/E$\\
\hline
Runtime& \tabincell{c}{$\mu^-$ mode 5 yrs+ $\mu^+$ mode 5 yrs}\\
\hline
Energy range & 100 MeV to 800 MeV  \\
\hline
Efficiency&  \tabincell{c}{ $\nu_{\mu}$ ($\bar{\nu}_{\mu}$) seclection: 50$\%$ \\ $\nu_e$ $(\bar{\nu}_e)$  selection: 40$\%$} \\
\hline
\tabincell{c}{Normalization\\error on signal} & \tabincell{c}{appearance channels: $2.5\%$\\ disappearance channels: 5$\%$}\\
\hline
\tabincell{c}{Normalization\\error on backgroud} & \tabincell{c}{ $5\%$ (all channels)}\\
\hline
\tabincell{c}{Backgound sources} & \tabincell{c}{Neutral current\\charge misidentifications\\atmospheric neutrinos}\\
\hline
\end{tabular}
\caption{Assumptions for near and far detectors in the simulation.}
\label{tab:glbtable}
\end{table}

\section{Physics performance of MOMENT}
\label{sec:results}
\subsection{Impacts on precision measurements of standard mixing parameters by CC-NSIs}
\begin{figure}[!t]
\includegraphics[scale=0.4]{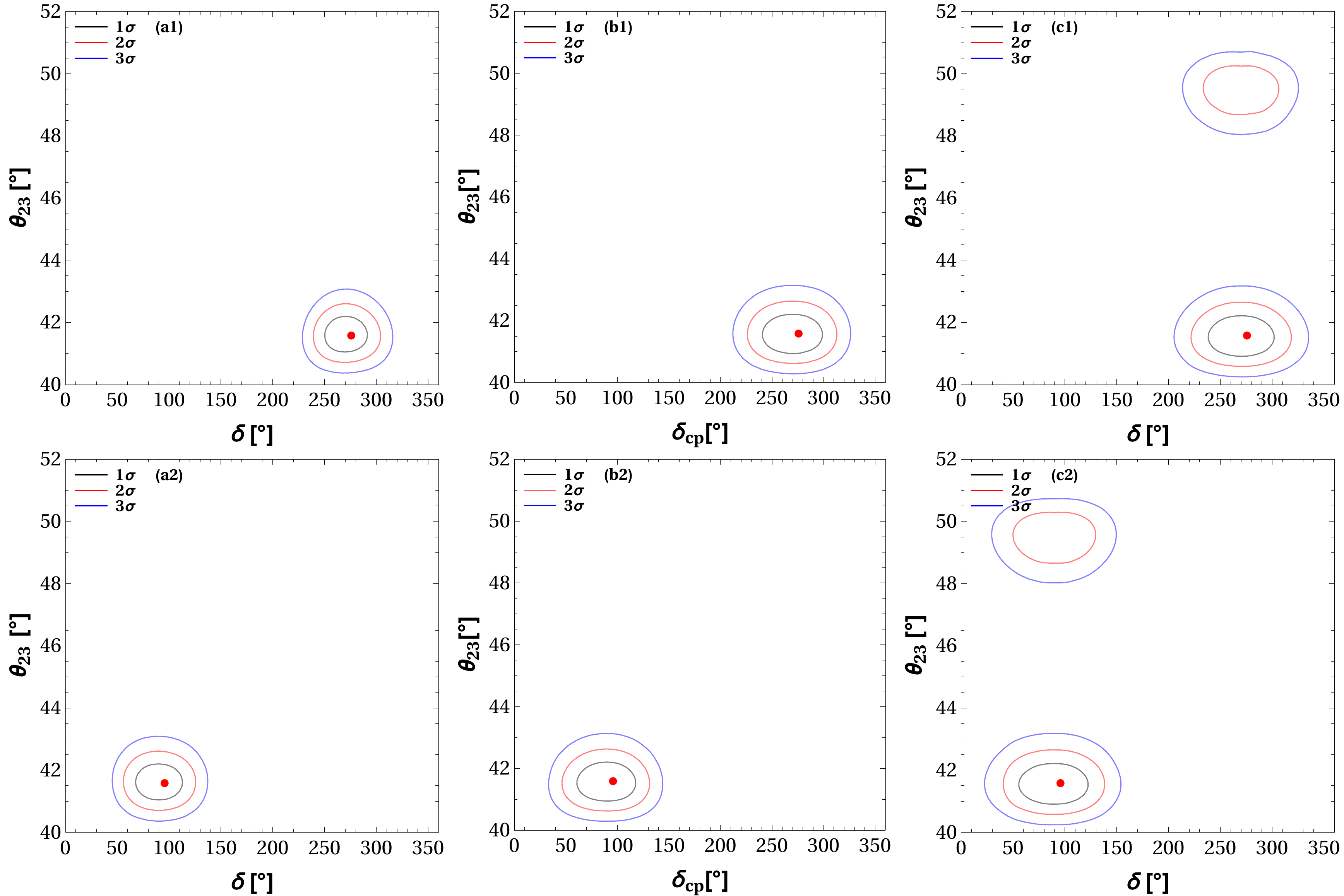}
\caption{The allowed region for $\theta_{23}-\delta$ for MOMENT. Panel (a) shows the determination of $\delta$ adn $\theta_{23}$ in the case of the standard three flavour frame.
In panel (b) we only consider the NSIs: $\epsilon_{e\mu}^s$, $\epsilon_{e\tau}^s$ and $\epsilon_{\tau e}^d$ which are related to $\nu_{\mu}$ appearance channels. (c) shows the effects 
of those NSIs related to the $\nu_{\mu}$ ($\bar{\nu}_{\mu}$) and $\nu_e$ ($\nu_e$) appearance channels: $\epsilon_{e\mu}^s$, $\epsilon_{e\mu}^d$, $\epsilon_{e\tau}^s$,
$\epsilon_{\mu e}^s$, $\epsilon_{\mu e}^d$, $\epsilon_{\tau e}^d$ and marginalization ranges are within current bounds given in \equ{bounds1} and \equ{bounds2}. All of the corresponding
phases can vary in (0, 2$\pi$). The red points in all panels indicate the true values.}
\label{fig:th23_cp}
\end{figure}
The CKM mixing matrix is well measured in the quark sector at the sub-percent level~\cite{Olive:2016xmw}, while mixing parameters in the lepton sector are far away from such a precision. It is very likely for the next-generation experiment like MOMENT to achieve the goal of doing precision measurements. In Fig.~\ref{fig:octant_prob}, we have showed that CC-NSIs may induce a bias in precision measurements of $\theta_{23}$ at the probability level. In this section, we take the Dirac CP-violating phase $\delta_{\textrm{cp}}$ and $\theta_{23}$ as an illustration to show the impacts from CC-NSIs after the simulation. The true value of $\theta_{23}$ is taken as 41.6$^{\circ}$. We choose two $\delta$ values with $\delta=\pi/2$ and $\delta=3\pi/2$ to simulate all oscillation channels at MOMENT and fit the neutrino spectra with/without NSIs. Fig. \ref{fig:th23_cp} demonstrates the numerical results. The true values of the standard oscillation parameters are shown by a red point in each panel. In all sub-figures (a), (b) and (c), we have considered uncertainties of standard mixing parameters. 
Panels (a1) and (a2) show the determination of $\delta_{\textrm{cp}}$ and $\theta_{23}$ in the case of the standard neutrino oscillation without NSIs. By running MOMENT, we can determine the mixing angle $\theta_{23}$ with an error bar of one degree at the $3\sigma$ confidence level, while the precision for $\delta_{\textrm{cp}}$ is good enough. In sub-figure (b) and (c), NSIs happening at the source and detector are turned on. All the corresponding CC-NSI phases can vary within $(0, 2\pi)$. In the panel (b1) and (b2), we only consider the CC-NSIs ($\epsilon_{e\mu}^s$, $\epsilon_{e\tau}^s$ and $\epsilon_{e\mu}^d$, and their marginalization ranges are allowed within the current bounds given in~\equ{bounds1} and~\equ{bounds2}.)
which are related to $\nu_{\mu}$ ($\bar{\nu}_{\mu}$) appearance channels. The panels of (b1) and (b2) show the enlarged uncertainties in parameter fittings. Especially, a degeneracy pops up in the measurement of $\theta_{23}$-$\delta_{\textrm{cp}}$ for the case of $\delta=3\pi/2$, while it is still safe for the case of $\delta=\pi/2$. Furthermore, we go to panels (c1)/(c2) by turning on those CC-NSIs related to the $\nu_{e}$ ($\bar{\nu}_{e}$) and $\nu_{\mu}$ ($\bar{\nu}_{\mu}$) appearance channels ($\epsilon_{e \mu}^s$, $\epsilon_{e \tau}^s$, $\epsilon_{e \mu}^d$, $\epsilon_{\mu e}^s$, $\epsilon_{\mu e}^d$, $\epsilon_{\tau e}^d$ and their marginalization ranges are within current bounds 
given in~\equ{bounds1} and~\equ{bounds2}). 

As we have discussed in Section~\ref{sec:oscP}, the $\nu_e$ appearance channel is affected by the parameters: $\epsilon_{\mu e}^s$, $\epsilon_{\mu e}^d$, $\epsilon_{\tau e}^d$, while the $\nu_{\mu}$
appearance channel is mainly determined by the parameters $\epsilon_{e\mu}^s$, $\epsilon_{e\tau}^s$, and $\epsilon_{e\mu}^d$.
This feature can be understood after a closer look at the \equ{far_pem_invacuum}. If $\delta$ is equal to $\pi/2$, the peak of the probability in $\nu_e\rightarrow\nu_{\mu}$ channel is much larger than the case  
of $\delta=3\pi/2$. In turn, the event rate in the detector for $\delta=\pi/2$ is much higher. The corresponding fitted results are much better. Therefore, CC-NSI parameters destroy the precise determinations of standard mixing parameters. As can be seen from panels (c1) and (c2), the degeneracy even shows up at high confidence levels when we consider all the relevant NSI 
parameters. We might get into the wrong best-fit region if we neglect the CC-NSIs from the new physics. A combination of different neutrino oscillation experiments might resolve such an ambiguity and finish the task of precision measurements of neutrino mixing parameters to the same level as quark mixing parameters. Or we need a more powerful machine, such as a neutrino factory.

\subsection{Correlations and constraints of NSI parameters}
\begin{figure}[!htbp]
\includegraphics[scale=0.5]{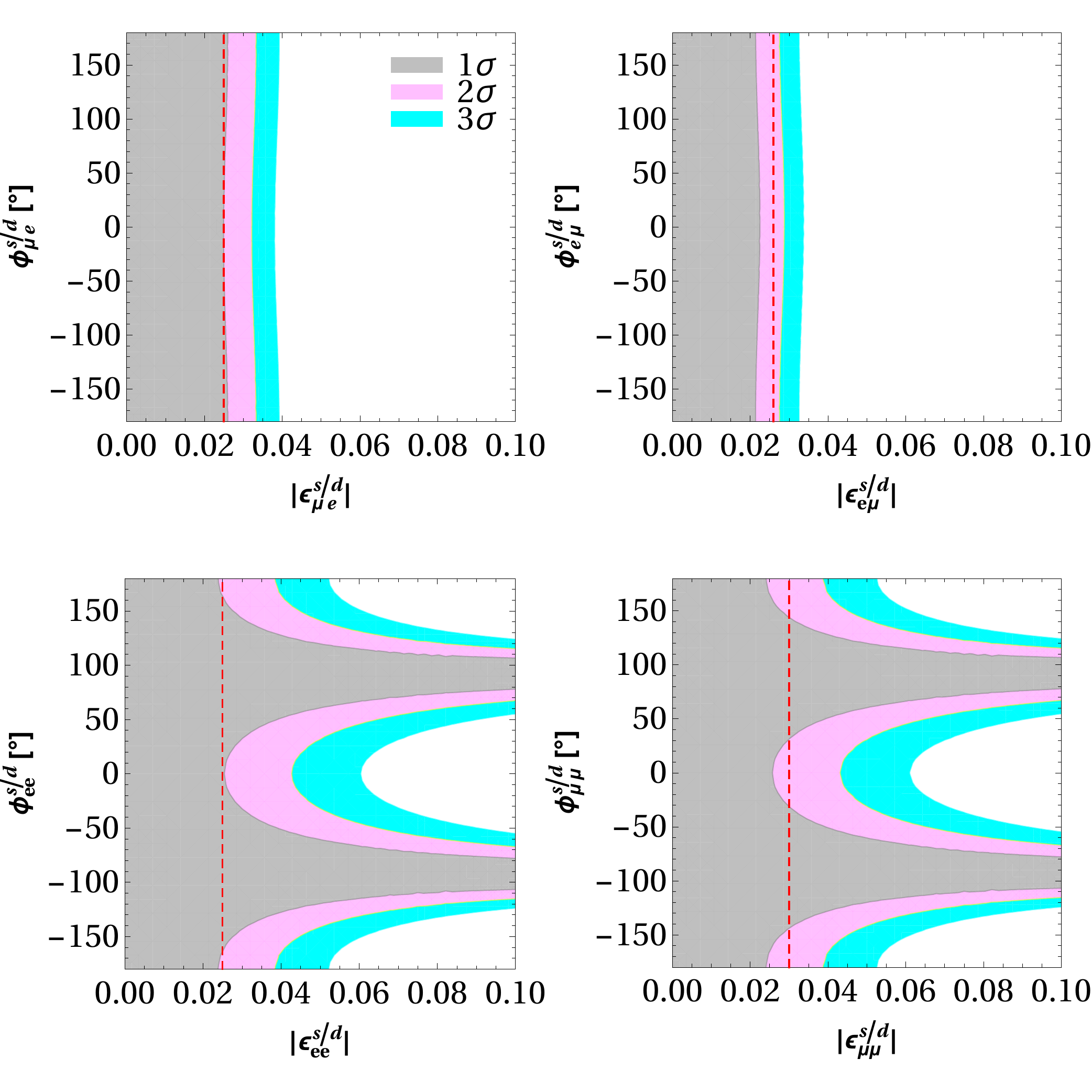}
\caption{Exclusion curves for the $|\epsilon_{\alpha\beta}^{s/d}|-\phi_{\alpha\beta}^{s/d}$ planes obtained by the near detector. The gray, magenta and cyan area 
is the allowed region at 1 $\sigma$, 2 $\sigma$ and 3 $\sigma$ C.L., respectively. The dashed line is the current bound.}
\label{fig:near_correlation_exclusion}
\end{figure}
\begin{table}[b]
\centering
\begin{tabular}{|c|c|c|c|c|}
\hline
Parameter & ND constraints & FD constraints& ND+FD constraints& Current bounds\\
\hline
\tabincell{c}{$|\epsilon_{ee}^s|$}& 0.028 & 0.029& 0.020 &0.025\\
\hline
\tabincell{c}{$|\epsilon_{e\mu}^s|$}& 0.024  &0.020& 0.017 &0.030 \\
\hline
\tabincell{c}{$|\epsilon_{e\tau}^s|$}& n/a &0.069& \textcolor{red}{0.069}&0.030\\
\hline
\tabincell{c}{$|\epsilon_{\mu e}^s|$}& 0.026 &0.023&0.018 &0.025\\
\hline
\tabincell{c}{$|\epsilon_{\mu\mu}^s|$}&  0.028  &0.030& 0.020 &0.030\\
\hline
\tabincell{c}{$|\epsilon_{\mu\tau}^s|$}& n/a  & 0.054 & \textcolor{red}{0.054}& 0.030\\
\hline
\tabincell{c}{$|\epsilon_{ee}^d|$}& 0.028  &0.027&0.019 &0.041 \\
\hline
\tabincell{c}{$|\epsilon_{e\mu}^d|$}&  0.024 & 0.017& 0.015 &0.026\\
\hline
\tabincell{c}{$|\epsilon_{\mu e}^d|$}&  0.026 & 0.024 &0.019 & 0.025 \\
\hline
\tabincell{c}{$|\epsilon_{\mu\mu}^d|$}&  0.028 & 0.030 & 0.020 &0.078\\
\hline
\tabincell{c}{$|\epsilon_{\tau e}^d|$}&   n/a  & 0.069 & \textcolor{red}{0.069} & 0.041\\
\hline
\tabincell{c}{$|\epsilon_{\tau\mu}^d|$}& n/a  &  0.054 & \textcolor{red}{0.054}&0.013\\
\hline
\end{tabular}
\caption{Expected $90\%$ credible regions on NSI parameters with a single detector or a combination of near and far detectors at the MOMENT experiment. Here NSI parameters are assumed to be real, or NSI-induced CP phases are switched off.}
\label{tab:boundtable}
\end{table}
\begin{figure}[!htbp]
\includegraphics[scale=0.53]{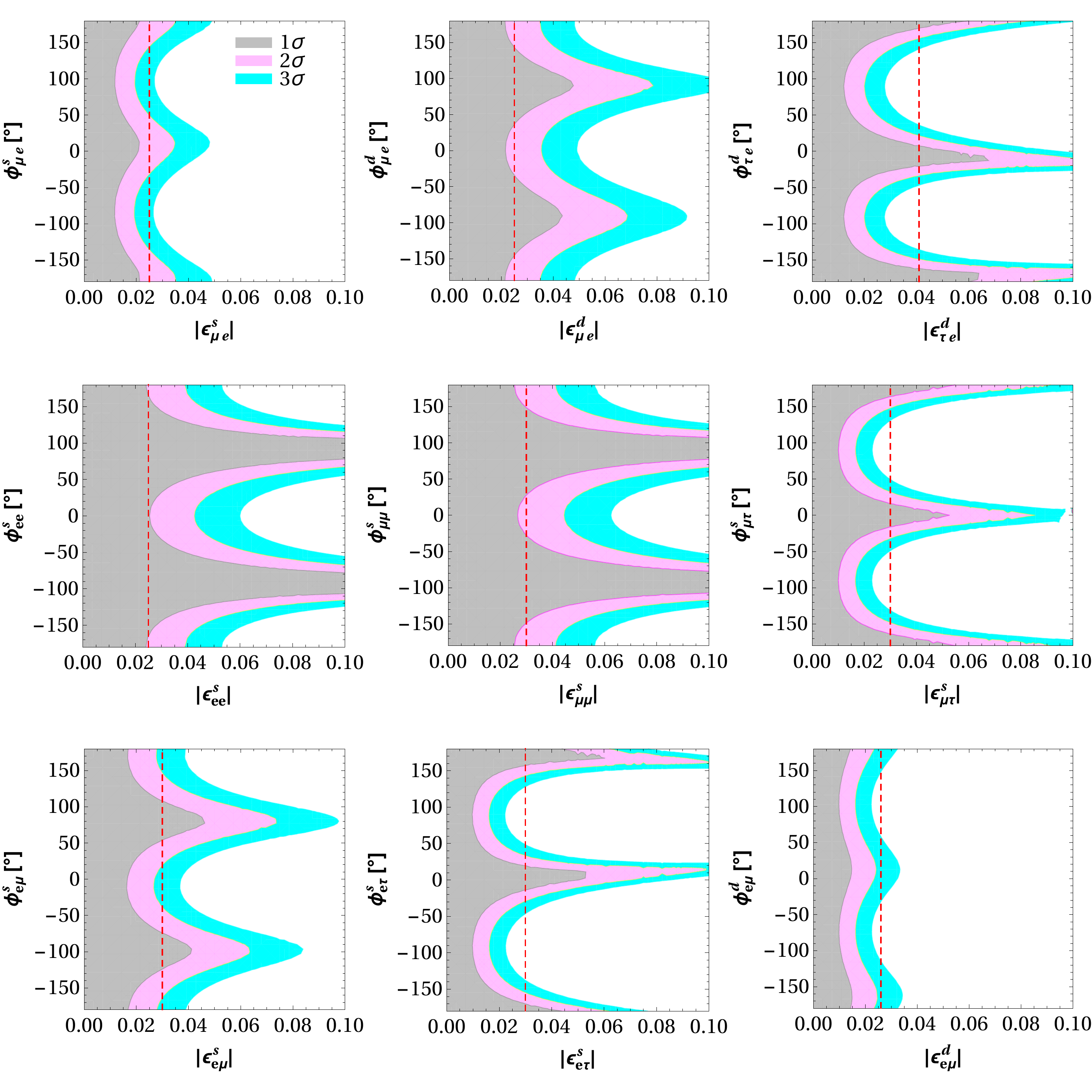}
\caption{Exclusion limits in the $|\epsilon_{\alpha\beta}^{s/d}|-\phi_{\alpha\beta}^{s/d}$ plane obtained from the far detector. The gray, magenta and cyan area 
is the allowed region at 1 $\sigma$, 2 $\sigma$ and 3 $\sigma$ C.L., respectively. The dashed line is the current bound.}
\label{fig:far_correlation_exclusion}
\end{figure}
We have introduced the NSIs by integrating the potential heavy propagator from the new physics scale based on the effective theory. Each NSI parameter has a magnitude which tells us the strength of new couplings and its associated phase to bridge the CP violating story. Therefore, it is convenient to adopt two methods: one is to take the NSI parameters as real, which corresponds to the strength of the coupling constant or switch off the NSI-induced CP violation phases; the other is to keep general assumptions given complex NSI parameters. In this section we discuss the constraints on source and detector NSIs from the far and near detector, respectively. For the former case, Table~\ref{tab:boundtable} demonstrates the sensitivity of MOMENT in constraining the NSI parameters using the single-parameter-fit at 90$\%$ C.L. The results are obtained by using all the neutrino and antineutrino oscillation channels with both near and far detectors. In a comparision of the current bounds and the expected limits from our simulation, we find that the MOMENT experiment with running time of 5+5 years has a potential to improve the constraints for the CC NSIs. In Table~\ref{tab:boundtable}, one can observe that most of the bounds for NSIs are improved except for $\epsilon_{e\tau}^s$, $\epsilon_{\mu\tau}^s$, $\epsilon_{\tau e}^d$ and $\epsilon_{\tau\mu}^d$ which are marked by red fonts. These results also confirm our observations in Section~\ref{sec:oscP}:
\begin{itemize}
 \item {NSI parameters $\epsilon_{ee}^{s/d}$ and $\epsilon_{\mu\mu}^{s/d}$, at the leading order of NSI terms in the disappearance channels can be well constrained with a combination of near and far detectors.} 
 \item {$\epsilon_{e\tau}^s$, $\epsilon_{\mu\tau}^s$, $\epsilon_{\tau e}^d$ and $\epsilon_{\tau\mu}^d$ cannot be well constrained by the near detector due to their negligible effects. Their constraints by the far detector are even weaker than the current bounds since their contributions to the dominant order of $\mathcal{O}(\epsilon s_{13})$ in \equ{far_pme_invacuum} and \equ{far_pem_invacuum} would be zero when $\delta=270$ degrees.} 
 \item {The constraints on $\epsilon_{e\mu}^s$, $\epsilon_{\mu e}^s$, $\epsilon_{e\mu}^d$, $\epsilon_{\mu e}^d$ are mainly from the appearance channels. A combination of near and far detectors have stronger constraints on $\epsilon_{e\mu}^s$ and $\epsilon_{e\mu}^d$ than $\epsilon_{\mu e}^s$ and $\epsilon_{\mu e}^d$. Fig.~\ref{fig:events} can be used to explain this property: the channel $\nu_e\rightarrow\nu_{\mu}$ has more events compared to the channel $\nu_{\mu}\rightarrow\nu_e$. In addition, it should be noted that $\nu_e$ and $\bar{\nu}_e$ disappearance channels are also sensitive to $\epsilon_{e\mu}^s$ and $\epsilon_{\mu e}^d$.}
\end{itemize}

In the following part, we will focus on the exclusion curve of the amplitude versus its corresponding phase for each NSI parameter. Based on the previous discussions in Section.~\ref{sec:oscP}, we neglect certain parameters which have trivial influence on the probabilities. We generate the event spectra with the true central values of standard oscillation parameters given in Table~\ref{tab:nu-fitvalues}. Then we turn on one NSI parameter and scan its amplitude and phase to fit the data. At the near detector, we pay more attention to these parameters: $\epsilon_{ee}^{s/d}$, $\epsilon_{\mu\mu}^{s/d}$, $\epsilon_{\mu e}^{s/d}$ and $\epsilon_{e\mu}^{s/d}$. At the far detector, parameters $\epsilon_{ee}^{s/d}$,$\epsilon_{\mu\mu}^{s/d}$, $\epsilon_{\mu\tau}^s$, $\epsilon_{\tau\mu}^d$, $\epsilon_{\mu e}^s$, $\epsilon_{\mu e}^d$, $\epsilon_{\tau e}^d$, $\epsilon_{e\mu}^s$, $\epsilon_{e\tau}^s$ and $\epsilon_{e\mu}^d$ are taken into account.
The results at a near/far detector are presented in
Fig.~\ref{fig:near_correlation_exclusion} and Fig.~\ref{fig:far_correlation_exclusion}, respectively.
In Fig.~\ref{fig:near_correlation_exclusion}, we show the excluded parameter space without any color at a near detector. One can observe that $\nu_{\mu}$ disappearance channel almost has the same performance with the $\nu_{e}$ disappearance channel. When $\phi_{ee}^{s/d}$ ($\phi_{\mu\mu}^{s/d}$) equals to 
$\pm\pi$ or zero, the corresponding amplitude $|\epsilon_{ee}^{s/d}|$ ($|\epsilon_{\mu\mu}^{s/d}|$) has the best limit. When these phases are equal to $\pm\pi/2$, the sensitivity to the amplitude disappears. For the $\nu_e$ ($\nu_{\mu})$
appearance channels, the constraint to $|\epsilon_{\mu e}^{s/d}|$ ($|\epsilon_{e\mu}^{s/d}|$) is almost irrelevant to the phase $\phi_{\mu e}^{s/d}$ ($\phi_{e\mu}^{s/d}$) supposing we only consider the source NSI parameter $\epsilon_{\mu e}^s$($\epsilon_{e \mu }^s$) or
detector NSI parameter $\epsilon_{\mu e}^d$($\epsilon_{e \mu }^d$). This is because the term containing the phases $\phi_{\mu e}^s$ and $\phi_{\mu e}^d$ plays an important role in two amplitudes as can be seen from~\equ{near_pme_antipme} and \equ{near_pem_antipem}. Appearance channels can well constrain the magnitude of related CC-NSI parameters while they barely have impacts on their phases. On the other hand, disappearance channels exclude a large parameter space allowed in the current experimental bounds. There is a strong correlation between the coupling strength and its phases in neutrino oscillation experiments.

In Fig.~\ref{fig:far_correlation_exclusion}, we switch to the sensitivities of CC-NSI parameters at a far detector. The colorful regions are allowed after we run a far detector at the MOMENT experiment. Compared to the current bounds marked by the dashed red lines, we obtain good constraints on most of NSI parameters at MOMENT, especially for $\epsilon_{\mu e}^s$ and $\epsilon_{e \mu }^d$. Here we list the features about the exclusion curves at the far detector for MOMENT:
\begin{itemize}
\item {The oscillation channels $\nu_{\mu}\rightarrow\nu_e$ ($\bar{\nu}_{\mu}\rightarrow\bar{\nu}_e$) and $\nu_e\rightarrow\nu_{\mu}$ ($\bar{\nu}_e\rightarrow\bar{\nu}_{\mu}$) are T-conjugate inverse of each other, leading to the symmetry 
between their NSIs: $\epsilon_{\mu e}^s$ and $\epsilon_{e\mu}^d$ have equal
contributions in the probability of $P^{FD}_{\nu_{\mu}\rightarrow\nu_e}$ and $P^{FD}_{\nu_e\rightarrow\nu_{\mu}}$, respectively. Similarly, the pair 
of $\epsilon_{\tau e}^d$ and $\epsilon_{e\tau}^s$ and the pair of $\epsilon_{\mu e}^d$ and $\epsilon_{e\mu}^s$ follow the same way. Therefore, these pairs of parameters have similar 
behaviour in Fig.~\ref{fig:far_correlation_exclusion}, which can be manifested by \equ{far_pme_2} and \equ{far_pem_2} in Section.~\ref{sec:oscP}.
The dependence of constraints for $|\epsilon_{\mu e}^s|$ ($|\epsilon_{e\mu}^d|$) on the corresponding phase is not strong, because $|\epsilon_{\mu e}^s|$ ($|\epsilon_{e\mu}^d|$) depends on both terms of $\sin(\delta+\phi_{\mu e}^s)$ (or $\sin(\delta-\phi_{e\mu}^d)$) and
$\cos(\delta+\phi_{\mu e}^s)$ (or $\cos(\delta-\phi_{e\mu}^d)$), which complement each other when varying the phase $\phi_{\mu e}^s$ ($\phi_{e\mu}^d$). Although $|\epsilon_{e\mu}^d|$ ($|\epsilon_{e\mu}^s|$) also depends on both terms of $\cos(\delta+\phi_{\mu e}^d)$ (or $\cos(\delta-\phi_{e\mu}^s)$) and
$\sin(\delta+\phi_{\mu e}^d)$ (or $\sin(\delta-\phi_{e\mu}^s)$), the former term is suppressed by the coefficient $c_{2\times23}$. Thus, the exclusion curve of $|\epsilon_{\mu e}^d|
-\phi_{\mu e}^d$ ($|\epsilon_{e\mu}^s|-\phi_{e\mu}^s$) mainly varies with the sine term.
However, the constraint on $|\epsilon_{\tau e}^d|$ ($|\epsilon_{e\tau}^s|$) only depends on the term of $\cos(\delta+\phi_{\tau e}^d)$
(or $\cos(\delta-\phi_{e\tau}^s)$) so that the limit to $|\epsilon_{\tau e}^d|$ ($|\epsilon_{e\tau}^s|$) will be 
extremely weak for $\phi_{\tau e}^d$ ($\phi_{e\tau}^s$)=0, $\pm\pi$ and be the best for $\phi_{\tau e}^d$ ($\phi_{e\tau}^s$)=$\pm\pi/2$.}
\item {Since we can take advantage of the $\nu_e$ ($\bar{\nu}_e$) channel in MOMENT, obtaining the sensitivities of $\epsilon_{e\tau}^s$, $\epsilon_{e\mu}^s$ and $\epsilon_{e\mu}^d$ would be accessible. It is noted that 
these parameters can not be constrained well in superbeam experiments.}
\item {The sensitivities to $\epsilon_{e\mu}^s$ and $\epsilon_{e\tau}^s$ are mainly extracted from the $\nu_\mu$ and $\bar{\nu}_\mu$ appearance channels. It should be noted, however, that the $\nu_e$
and $\bar{\nu}_e$ disappearance channels can help to enhance the constraints of them. Similarly, with contributions from $\nu_e$
and $\bar{\nu}_e$ disappearance channels, the sensitivities to $\epsilon_{\mu e}^d$ and $\epsilon_{\tau e}^d$ will be improved.}
\item {The constraints to NSI parameters $\epsilon_{ee}^s$, $\epsilon_{\mu\mu}^s$, $\epsilon_{\mu\tau}^s$, $\epsilon_{\tau\mu}^d$ are extracted from the disappearance channels. The exclusion curves for $\epsilon_{ee}^{s/d}$ and $\epsilon_{\mu\mu}^{s/d}$ are similar to each other at the near detector, since they are entangled with the same term in oscillation probabilities.
In addition, there are some symmetric relationships between the source and detector NSI parameters, such as the pair of $\epsilon_{\mu\tau}^s$ and $\epsilon_{\tau\mu}^d$, the pair of $\epsilon_{ee}^s$ and $\epsilon_{ee}^d$, the pair of $\epsilon_{\mu\mu}^s$ and $\epsilon_{\mu\mu}^d$. This is not surprising at all, since the pair of effective coupling constants will be the same for neutrinos produced and detected related to the same charged leptons. }
\end{itemize}
The far detector has a good sensitivity to NSI parameters, especially for $\epsilon_{e\mu}^d$ and $\epsilon_{\mu e}^s$. The numerical results of the correlations between the amplitudes and phases can be interpreted with the previous probability-level discussions. Almost all NSI-induced phases change the exclusion limits severely except the e-mu sector. Meanwhile, limits on other sectors are not as good as those on the e-mu stamped CC-NSIs. Therefore, MOMENT using muon-decay beams has its unique capability of improving the constraints on $\epsilon_{e\mu}^d$ and $\epsilon_{\mu e}^s$.

\section{Summary and conclusions}
\label{sec:summary}
New Physics beyond SM might cause non-standard neutrino interactions and leave imprints on the neutrino oscillation. The next-generation accelerator neutrino experiment MOMENT intends to produce the powerful neutrino beam with an energy of O(100) MeV by muon decays and leaves plenty of room for detector selections and physics study. At this energy range, quasi-elastic neutrino interactions dominate the detection process and backgrounds from $\pi^0$ are highly suppressed. Compared with traditional superbeams from charged meson decays where intrinsic backgrounds have to be alleviated by the off-axis technology like T2K and NO$\nu$A, beams from muon decays are cleaner neutrino sources and good at a detection of new physics. CC-NSIs happening at neutrino productions and detections point to the new phenomenon, where a neutrino produced or detected together with the charged lepton will not necessarily share the same flavour, and flavour conversion is present already at the interaction level and ``oscillations" can occur at zero distance. With the capability of flavour and charge identifications, we have an opportunity to use eight appearance and disappearance oscillation channels in the physics study. We have chosen the advanced neutrino detector using the Gd-doped Water Cherenkov technology and studied neutrino oscillations confronting with CC-NSIs at the MOMENT experiment. 

In order to understand the relevant behavior from NSIs, we have perturbatively derived oscillation probabilities including CC-NSIs at a short and far distance, and tried to analyze parameter correlations of standard neutrino mixing parameters with NSI parameters. We have investigated impacts of the charged current NSIs at the neutrino oscillation probabilities, selected the following dominanting CC-NSI parameters $\epsilon_{ee}^{s/d}$, $\epsilon_{\mu\mu}^{s/d}$, $\epsilon_{\mu e}^{s/d}$ and $\epsilon_{e\mu}^{s/d}$ for the near detector, and concentrated on $\epsilon_{ee}^{s/d}$, $\epsilon_{\mu\mu}^{s/d}$, $\epsilon_{\mu\tau}^s$, $\epsilon_{\tau\mu}^d$, $\epsilon_{\mu e}^s$, $\epsilon_{\mu e}^d$, $\epsilon_{\tau e}^d$, $\epsilon_{e\mu}^s$, $\epsilon_{e\tau}^s$ and $\epsilon_{e\mu}^d$ for the far detector. A near detector at MOMENT is good at detecting the zero-distance effects induced by NSIs while the oscillation pattern would have not been developed in the standard neutrino oscillation paradigm. With near and far detectors, we have found that CC-NSIs can induce bias in precision measurements of standard mixing parameters. Taking $\delta_{cp}$ and $\theta_{23}$ as an example, we have found degeneracies after introducing CC-NSI parameters. With a non-maximal $\theta_{23}$, its degeneracy with the standard CP phase $\delta_{\mathrm{CP}}$ gets much worse if CC-NSIs appear in the neutrino production and detection processes. The current bounds on NSI parameters governing the neutrino productions and detections are about one order of magnitude stronger than those related to neutrino propagation in matter. Our study has shown that a combination of near and far detectors at MOMENT is able to provide lower bounds on CC-NSIs where a factor of about two can be envisaged for most of parameters compared with the current experimental bounds, as shown in Table~\ref{tab:boundtable}. We have found strong correlations of NSIs and constrained NSI parameters using a combination of near and far detectors at MOMENT.

The feasibility of physics performance at MOMENT strongly depends on inputs of the accelerator facility and the advanced neutrino detection technology. In the future, results will be further improved by tuning the beam energy and optimizing the baseline. We hope that our study will boost the the research and development activities for MOMENT.

\section{Acknowledgement}
This work is supported by the start-up funding from SYSU, the National Natural Science Foundation of China under Grant No. 11505301 and the Special Program for Applied Research on Super Computation of the NSFC-Guangdong Joint Fund (the second phase) under Grant No.U1501501. YBZ appreciates valuable discussions with Steven Wong and Amir Khan. We thank Jiajun Liao and David Vanegas Forero's suggestions and comments on the preliminary draft. We would like to thank the accelerator working group of MOMENT for fruitful discussions concerned with neutrino fluxes. Furthermore, we are grateful to IHEP colleagues for collaborations, especially Jingyu Tang, Yu-Feng Li, Miao He and Nikos Vassilopoulos. Dr. Neill Raper is highly appreciated for his patient proofreading of our manuscript.

\bibliography{references.bib}

\end{document}